\pdfoutput=1

\documentclass[twoside,leqno,twocolumn]{article}

\usepackage[letterpaper]{geometry}

\usepackage{ltexpprt}
\usepackage{graphicx}
\usepackage{biblatex}
\usepackage{booktabs}
\usepackage[ruled,vlined, algo2e]{algorithm2e}
\usepackage{siunitx}
\usepackage{hyperref}

\usepackage{amsmath, amsfonts, amssymb, amsthm}

\hypersetup{colorlinks,linkcolor=blue,urlcolor=blue, citecolor=}

\newtheorem{thm}{Theorem}
\DeclareMathOperator*{\argmin}{argmin} 

\begin{document}

\title{\Large Fairmandering: A column generation heuristic for \\fairness-optimized political districting\thanks{Supported in part by NSF grants CCF-1522054, CCF-1526067, and CCF-1740822, DMS-1839346, and CNS-1952063.}
}
 \author{Wes Gurnee\thanks{Cornell University}
 \and David B. Shmoys\thanks{Cornell University}}

\date{}

\maketitle






\begin{abstract} \small\baselineskip=9pt 
The American winner-take-all congressional district system empowers politicians to engineer electoral outcomes by manipulating district boundaries. Existing computational solutions mostly focus on drawing unbiased maps by ignoring political and demographic input, and instead simply optimize for compactness. We claim that this is a flawed approach because compactness and fairness are orthogonal qualities, and introduce a scalable two-stage method to explicitly optimize for arbitrary piecewise-linear definitions of fairness. The first stage is a randomized divide-and-conquer column generation heuristic which produces an exponential number of distinct district plans by exploiting the compositional structure of graph partitioning problems. This district ensemble forms the input to a master selection problem to choose the districts to include in the final plan. Our decoupled design allows for unprecedented flexibility in defining fairness-aligned objective functions. The pipeline is arbitrarily parallelizable, is flexible to support additional redistricting constraints, and can be applied to a wide array of other regionalization problems. In the largest ever ensemble study of congressional districts, we use our method to understand the range of possible expected outcomes and the implications of this range on potential definitions of fairness.
\end{abstract}

\section{Introduction}

Section 4, Article 1: ``The Times, Places and Manner of holding Elections for Senators and Representatives, shall be prescribed in each State by the Legislature thereof; but the Congress may at any time by Law make or alter such Regulations.'' Other than this brief statement regarding regulatory authority, the Constitution offers little guidance on the process by which the members of the House of Representatives are elected. Our current district-based system, where districts are most often drawn by state legislatures, enable self-interested and self-preserving politicians to secure a partisan advantage, suppress the vote of minority groups, and protect incumbents from competition. Such practices, broadly known as gerrymandering, enable large discrepancies between the popular vote and political representation, and ensure that electoral outcomes are unresponsive to swings in public opinion. In 2021, 428 congressional, 1938 state senate, and 4826 state house districts will be redrawn, cementing the power balance in the United States for the next decade.

Given the mathematical richness and societal importance of redistricting, there is a large literature bridging computational and political sciences on the causes \cite{erikson1972malapportionment, stephanopoulos2017causes}, consequences \cite{gelman1990estimating, mccarty2009does}, and potential remedies of gerrymandering \cite{polsby1991third, berman2004managing, vickrey1961prevention}. Until recently, nearly all proposed computational solutions to create more equitable maps focused on creating maximally compact maps drawn without any political data as input \cite{vickrey1961prevention, hess1965nonpartisan, garfinkel1970optimal, nygreen1988european, mehrotra1998optimization, li2007quadratic, ricca2008weighted, ricca2013political, cohen2018balanced, levin2019automated}. This is in part due to the computational hardness of optimizing for fairness at scale; it is also due to the difficulty of acquiring granular and historical political data, and because some states explicitly ban the use of political data in the redistricting process \cite{levitt2008citizen}. While it is true that compactness-optimized maps are blind to partisan bias, it is not true these maps are free from partisan advantage. The two most successful lines of work that begin to address this gap in the literature are multi-level optimization algorithms \cite{swamymulti} and ensemble methods, most notably recombination Markov chains \cite{deford2019recombination}. 

Exact optimization methods are intractable for all but the smallest problems because optimal political districting is $NP$-hard for any useful objective function \cite{puppe2008computational, kueng2019fair, chatterjee2019partisan}. Optimization approaches for fairness therefore either rely on local search techniques \cite{ricca2008local, king2015efficient} or embed a heuristic in some stage of the optimization. The most successful formulation thus far does both by iteratively coarsening the problem input, solving to optimality for multiple objectives, and then uncoarsening with a local search routine at each level of granularity \cite{swamymulti}. However, the number of districts in the problem instance limits the achievable coarsening and therefore such a scheme is unable to scale to large states or state legislative boundaries where there are a much greater number of seats.

Ensemble methods \cite{liu2016pear, chen2013unintentional, deford2019recombination} generate a large ensemble of random maps to infer a property about the distribution of all feasible district plans. These methods are ideally suited for detecting gerrymandering \cite{herschlag2017evaluating, duchin2018outlier, chikina2017assessing} via outlier analysis with respect to the random ensemble of partisan blind plans and evaluating the effect of proposed rules \cite{cain2017reasonable} on the space of legal plans. However, because these techniques generate one map at a time (and are not optimizing for any property), they are not efficient at sampling from the tails of any distribution. This is problematic because extreme solutions, such as the most unfair solution under a new set of rules or the fairest possible map which maximizes other desirable qualities, are often the solutions we are looking for.

Our work is novel in that it combines these strands of literature and inherits the strengths of both. We introduce a new algorithm which creates an exponential ensemble of distinct district plans by exploiting the natural combinatorial compositionality of redistricting by generating an expressive set of districts that is also efficient to optimize over. The ensemble is exponential in $k$, the number of seats in a plan, making our approach ideally suited for larger states and state assembly districts. While these district plans are not appropriate for assessing certain statistical relationships given their lack of independence, they are nonetheless useful for understanding the range of possibilities that a state's geography and rules allow for. Additionally, by decoupling the generation of districts from the optimization of districts, our formulation enables a more flexible range of objective functions that we can use to optimize for arbitrary mappings from votes to seats, supporting a large family of fairness objectives.

Our core claim is that mapmakers should explicitly optimize for fairness rather than using questionable proxies like compactness. The main contribution of this paper is to introduce a new methodology to achieve this, while being extensible to a wide variety of other regionalization problems \cite{duque2007supervised}, particularly those where more complicated objectives outside of compactness are important. Our approach has a number of useful properties including parallelizability and scalability, making huge problem instances tractable, reusability of generated districts and flexibility of objectives, enabling rapid iteration of many definitions of fairness, as well as a modular and transparent execution that facilitates certain human-in-the-loop interactions. These properties allow us to solve problems at unprecedented scale and immediately provide value to independent commissions and other redistricting authorities to quickly and cheaply generate a large volume of high quality plans. In addition to detailed empirical results of different parameter configurations of our algorithm, we use its ensembling capabilities to perform the largest ever ensemble study to understand the range of possible expected electoral outcomes of all multi-district states and the implications of this range on potential definitions of fairness.

\section{Methodology}
\subsection{Preliminaries}
\begin{figure}[b]
    \centering
    \includegraphics[width=\linewidth]{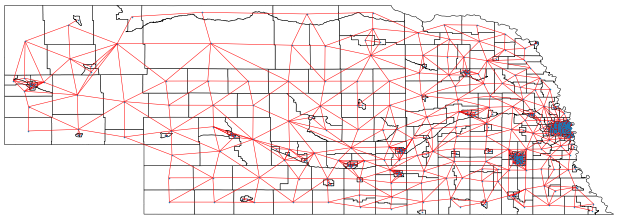}
    \caption{Adjacency graph of Nebraska's 532 census tracts.}
    \label{fig:Adj_graph}
\end{figure}

\paragraph{The Political Districting Problem (PDP)} 
The basic input to the PDP is a set of geographic polygons and their respective populations. Each polygon is represented as a node in a planar graph $G = (B, E)$ where $B$ is the set of atomic geographic blocks used to construct the districts and $(i, j) \in E$ whenever $b_i$ and $b_j$ are geographically adjacent. We let $p_i$ represent the population of $b_i$ and let $d_{ij}$ be the Euclidean distance between the geographic centroids of $b_i$ and $b_j$. In this setting, the atomic blocks are typically precincts, census tracts, census blocks, or an aggregation thereof. See Figure~\ref{fig:Adj_graph} for an example.

A feasible solution to the PDP for a state with $k$ districts is a $k-$partition of $B = D_1 \cup \dots \cup D_k$ (and $D_i \cap D_j = \emptyset$, for each pair $i,j \in \{1,\ldots,k\}, i\neq j$) such that each district $D_j$ is population balanced, geographically contiguous, and compact. That is, each district $D_i$ must have roughly equal population, the subgraph of $G$ with nodes in $D_i$ must be one fully connected component, and the district should have a geographic shape resembling a fairly simple convex polygon. A solution is called a districting or a district plan. Part of the challenge of this application domain is that these constraints are usually under-specified, with each state offering different levels of specificity and few national standards to fill the gaps. For Congressional districts, the Supreme Court ruled that the districts must be balanced "as nearly as is practicable" \cite{1964wesberry} but state legislature districts only require “substantial equality of population”\cite{1964reynolds}.  In practice this means adding the constraint that for each district $D_j$
$$\hat{p} (1 - \epsilon_p) \leq \sum_{i \in D_j} p_i \leq  \hat{p} (1 + \epsilon_p), \ \ \ \  j = 1, \dots, k$$
where $\hat{p} = \sum p_i / k$ is the ideal district population, with $\epsilon_p < .01$ for Congressional districts, and $\epsilon_p < .05$ for state legislatures.

There are many proposed mathematical measures of compactness in the redistricting literature \cite{young1988measuring}, but each has particular shortcomings, and none are universally agreed upon. Contiguity is also not entirely unambiguous, because of water features and states that have multiple connected components (e.g., Michigan). Any ambiguity must eventually be resolved in the adjacency graph $G$ by the practitioner.

\paragraph{Column Generation}
Most integer programming formulations for solving the PDP, and regionalization methods more broadly, choose a set of blocks that act as centers of a region and assign every block to exactly one center, all in one step. They associate decision variables for each pair of nodes \cite{ricca2013political} for the assignment of blocks to centers and an additional variable for each block to indicate whether or not it is a center. In this design paradigm, the generation of districts and the optimization are performed jointly.

Our approach is based on the idea that we can decouple the district generation from optimization by enumerating feasible districts as an input to the final optimization problem \cite{garfinkel1970optimal, nygreen1988european, mehrotra1998optimization}. An optimal solution could theoretically be found by enumerating every legal district $D_1,\ldots,D_N$ and solving a set partitioning problem to select the optimal $k$ districts using binary decision variables \cite{garfinkel1970optimal}. Formally, if $A=(a_{ij})$ is the block-district matrix where 
$$a_{ij} = \begin{cases} 1 \quad \text{if } b_i \in D_j \quad (\text{block } i \text{ is in district } j)\\
0 \quad \text{else}
\end{cases}$$
and $x$ is a binary decision vector where $x_j$ indicates whether or not $D_j$ is in the final plan, the set of feasible solutions is 
$$F = \{x: Ax = 1, \ \  ||x||_1 = k, \ \ x \in \mathbb{Z}_2^{N} \}.$$
An optimal plan corresponds to the $\hat{x} \in F$ that minimizes (or maximizes) some linear objective $cx$.
While viable for small problem sizes ($|B| < 100$), the number of feasible districts grows exponentially and quickly becomes intractable to enumerate. However, every feasible district plan must have exactly $k$ non-zero variables, implying that the vast majority of the column space is not useful, and therefore unnecessary to generate.

Implicit column generation is a well-studied technique to efficiently solve huge set partitioning problems \cite{barnhart1998branch} such as the PDP. These algorithms involve iteratively solving a relaxed master problem and transferring dual information to a column generation subproblem to determine if there are columns with negative reduced costs that should be included in the master problem \cite{mehrotra1998optimization}. However, we found when we scaled this approach to larger problem instances, the degeneracy of the master problem rendered the dual values meaningless, resulting in the generation of low quality columns. The core issue with traditional one-at-a-time column generation is that a single arbitrary district is unlikely to have $k-1$ other districts with which it can be composed to exactly cover the whole state. This observation motivates our heuristic that generates explicitly complimentary columns \cite{ghoniem2009complementary} that can be composed to express a huge number of legal plans. 

In addition to the degeneracy of the master problem, there are a number of features of fairness-optimized political redistricting that make ``optimal'' column generation the wrong approach. When optimizing for expected electoral outcomes, there is significant uncertainty in the objective function. Furthermore, there are many desirable and undesirable properties of feasible maps that should be balanced and analyzed in aggregate. Lastly, the difficulty of solving the master set partitioning problem scales with the number of districts (columns); however, depending on the construction of the districts, the feasible set of plans $F$, can scale much faster than the set of districts $D$. For these reasons, instead of focusing on certifying that our columns are  ``optimal,'' a better goal is to generate \textit{efficient} columns, those that maximize $|F| / |D|$.

\subsection{Stochastic Hierarchical Partitioning (SHP)}

\paragraph{Sample Tree}
What is the fastest way to generate one trillion unique district plans for a state with 16 districts? One way is to first partition the state into two and generate one million district plans for each half each with 8 districts. The fastest way to generate two sets of one million district plans? Split each half into quarters and generate one thousand district plans for each quarter. And to generate a thousand district plans for each quarter? The attentive reader guesses generating 32 plans for each eighth; partition each eighth 32 times and return.

The above procedure admits over one trillion district plans while generating only 512 districts by solving only 263 partitioning problems. In practice, none of the trillion plans might be any good because they are all similar, in that they respect the structure imposed by nodes higher in the tree. Therefore, in our method, rather than select a single split, we sample each level repeatedly; this exponential leverage of composing hierarchical partitions is the key property we exploit to generate efficient columns.
\begin{figure}[h]
    \centering
    \includegraphics[width=\linewidth]{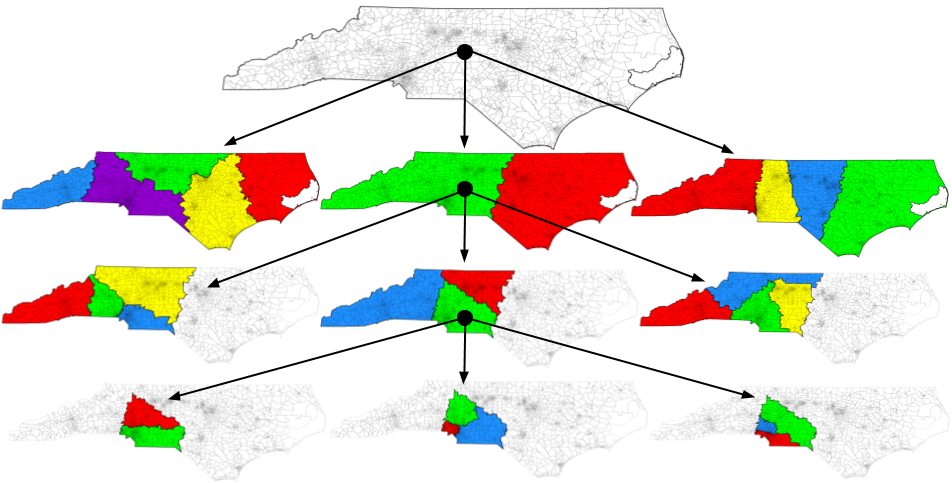}
    \caption{One branch of a sample tree with fan-out width $w=3$ for North Carolina.}
    \label{fig:tree}
\end{figure}

To formalize this notion, we introduce the concept of a SHP sample tree, a tree of compact and contiguous regions that maintain a compositional compatibility invariant (see Figure~\ref{fig:tree} for an example branch of a sample tree). A sample tree node ($R, s$) is a collection of contiguous blocks $R \subseteq B$ such that the population of region $R$ is sufficient for the construction of $s$ districts that respect the population constraint. We call $s$ the \textit{capacity} of the region. The root node of the sample tree is ($B, k$). The compatibility invariant requires that for any node ($R, s$), a valid district plan with $s$ districts can be constructed by combining any valid district plan from each of its children from a single sampled partition. Leaf nodes have $s = 1$; they constitute the set of generated districts and correspond to the columns of the block-district matrix $A$.

The shape of the tree dictates the properties of the column space. The two main parameters that we use to control the shape are the sample fan-out $w$idth $w$, and the split si$z$e $z$. The former is the number of partitions we sample at each node, and the latter is the size of the partition (binary, ternary, etc.). Parameters $w$ and $z$ control the inherent trade-off in the efficiency and diversity of the columns generated. Sampling more at higher nodes in the tree increases diversity but decreases efficiency. Sampling more at lower nodes has the opposite effect. Increasing the average split size $z$ leads to a shorter tree which yields less exponential leverage but more diversity.

We build the tree by sampling multiple random partitions of each node starting with the root. For computational reasons, we implement this iteratively, rather than recursively, using a queue initialized with the root ($B, k$). We continue to partition regions into smaller regions and add those regions to the queue unless $s=1$, in which case we add this region to the set of generated districts. The \textit{root partition} imposes significant structure on the remaining partitions, so in practice, for sake of diversity, we typically sample hundreds or thousands of root partitions; in contrast, each nonleaf child node is sampled only a few times (between 2 and 10). The problem then becomes creating an expressive enough column space by sampling a diverse enough set of region partitions.

\begin{figure}
    \centering
    \includegraphics[width=\linewidth]{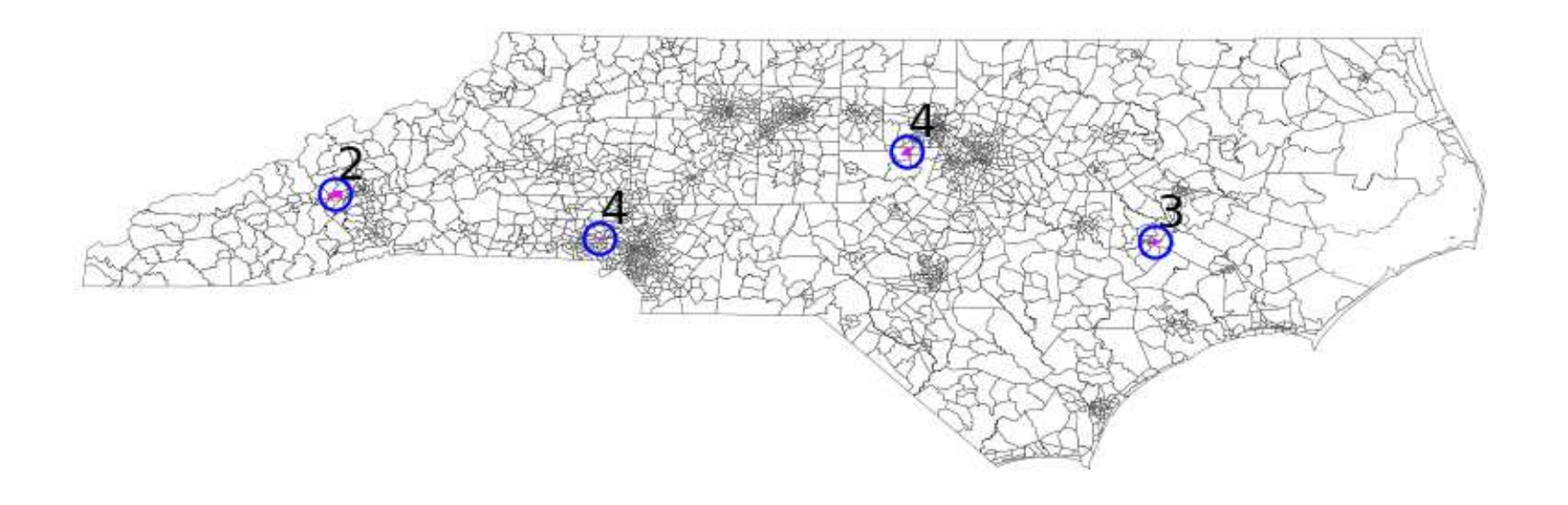}
    \includegraphics[width=\linewidth]{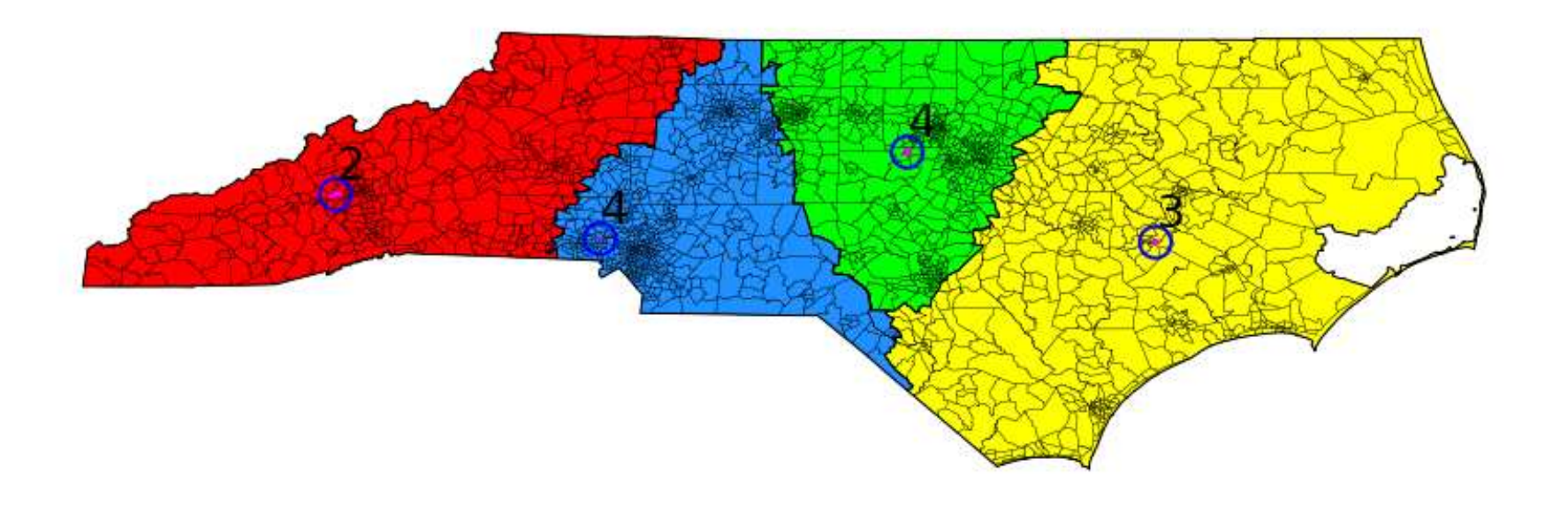}
    \caption{An example partition step of the sample tree root node with split size $z = 4$. First, we sample center population capacities and their geographic locations. Second, we solve the partition integer program to partition the region into contiguous subregions of appropriate population.}
    \label{fig:partition}
\end{figure}

\paragraph{Partition}
Each random partition of a node follows multiple steps: sample the split size, sample the capacity of each center, sample the position of the centers, and finally solve the partitioning integer program to perform the final assignment of blocks to regions (see Figure~\ref{fig:partition}).  

In more detail, for a node $(R, s)$, encoding a region $R\subseteq B$ with capacity for $s$ districts, we first sample the split size $z \in [z_{min}, \min(s, z_{max})]$ corresponding to the target number of children and then uniformly at random sample the capacity of each child node $s_1, \dots, s_z$ such that $\sum_i s_i = s$. To prevent significant imbalance in the tree, we usually add a constraint on the maximum difference between the maximum and minimum subregion capacity. One could imagine also tuning $z_{min}$ and $z_{max}$ depending on the region capacity $s$ and the overall size of the state $k$, but for simplicity, in our experiments we simply set $z_{min}=2$ and $z_{max}=5$; this works well in practice.

The next step is to sample the set of centers $C \subset R$ that will act as the seeds for each region of the partition. We detail a number of ways we do this in Appendix~\ref{algo_details}, but the goal is to let the randomization enable a selection of diverse centers that are well-spaced enough to not sacrifice feasibility or create contorted and noncompact regions. Each center is then matched to a size $s_i$ based on an idealized capacity given by the region's Voronoi diagram with respect to the set of centers. See Appendices~\ref{algo_details} and \ref{generation_experiment} for explanations of different center selection and capacity matching methods with detailed empirical results for each proposed method. 

Finally, using these centers and capacities, we solve an integer program to assign blocks to the $z$ subregions, each centered at a center $c_i$ with capacity $s_i$ that maximize compactness and satisfy the population balance and contiguity constraints. Enforcing contiguity is one of the reasons political districting problems are so difficult, since it requires an exponential number of constraints to do exactly. Instead of explicitly enforcing contiguity, we use an approach first presented in \cite{mehrotra1998optimization} that enforces each partition be a sub-tree of the shortest path tree rooted at the district center. This requires computing the shortest path distance from each center to all nodes in the subgraph of $G$ with nodes in region $R$, denoted $G_R$ with edgeset $E_R$. Let $d^R_{ij}$ be the shortest path distance between $b_i$ and $b_j$ in the subgraph $G_R$. For a center $i$ and block $j$, let 
 $$S_{ij} = \{k : \ (k, j) \in E_R, \ d^R_{ik} < d^R_{ij} \};$$ for a block $j$ to be assigned to a center $i$, then one element of $S_{ij}$, the set of neighbors of $j$ closer to $i$, must also be assigned to center $i$. 

Formally, the partition integer program (PIP) with inputs $R$ and $C$ is as follows:
\setcounter{equation}{0}
\begin{align}
\text{minimize}\quad  & \displaystyle\sum\limits_{i \in C} \displaystyle\sum\limits_{j \in R} (d_{ij})^\alpha p_j x_{ij} &\\
\text{s.t.} \quad& \displaystyle\sum\limits_{i \in C}  x_{ij} = 1,  & \forall j  \in R \\
&\displaystyle\sum\limits_{j \in R} p_j x_{ij} \leq  \hat{p} (s_i + \epsilon_p) & \forall i \in C \\
&\displaystyle\sum\limits_{j \in R} p_j x_{ij} \geq  \hat{p} (s_i - \epsilon_p)  & \forall i \in C \\
& \displaystyle\sum\limits_{k \in S_{ij}}  x_{ik} \geq x_{ij},  &\forall i \in C , \quad \forall j \in R \\
                 &                                                x_{ij} \in \{0,1\}, &\forall i \in C , \quad \forall j \in R
\end{align}
where constraint (2.2) enforces that each block is assigned to exactly one subregion, constraints (2.3) and (2.4) enforce population balance, and constraint (2.5) enforces contiguity. The objective function is based on the dispersion, or moment of inertia measure of compactness \cite{young1988measuring}. We apply a cost exponential to the distance term drawn uniformly at random $\alpha \in [1, 2]$ to promote more diverse region shapes. For states with abnormal geometric features (e.g. those with multiple connected components or with highly nonconvex shapes such as NY, MA, and MI), it is preferable to optimize for network distance rather than Euclidean distance. We choose this method of enforcing contiguity because it is efficiently computable, but it is not without its drawbacks. While it prevents the formation of noncompact districts, it also discounts many districts that would pass an eyeball test and generally reduces the space of feasible solutions. One item of future work is to profile different methods of enforcing contiguity \cite{validi2020imposing}.

The capacitated partition program shares many similarities with the capacitated transportation problem, which is solvable in polynomial time with flow networks \cite{ford1956solving}. However, instead of assigning units of supply to units of demand, we are assigning units of population to districts with population requirements. We include the additional contiguity constraint [5], but given that most compact solutions are contiguous, this does not drastically increase the difficulty. For a region with $n$ nodes and $z$ centers, the partition IP has $nz$ variables and $n + 2z + nz$ constraints. These problems rarely take more than a few seconds to construct and solve.

\paragraph{Tree Complexity}
 The first key property of our approach is the leverage we gain from our tree formulation: districts with shared ancestors are always compatible. With a deeper tree, the number of feasible district plans increases rapidly. Importantly, this makes our method more effective when $k$ is larger, whereas most optimization approaches become intractable with large $k$.

Let $P(R, s)$ denote the number of feasible $s$-district plans of region $R$ given by the sample tree. If $w(R)$ gives the sample width for region $R$, and $z(i)$ gives the number of subregions for sample $i$, the number of plans has a recursive formula
$$ P(R, s) = \begin{cases}
\sum_{i=1}^{w(R)} \prod_{j=1}^{z(i)} P(R_{ij}, s_j)\ & \text{for} \ \ s > 1 \\
1 & \text{for} \ \ s = 1
\end{cases}$$
where $R_{ij}$ is subregion $j$ of sample $i$. Intuitively this means summing over sibling samples and multiplying over children. 

\begin{thm}
Consider a set of blocks $B$ to be partitioned into $k$ districts. For a sample tree with root node $(B, k)$ and with nodes corresponding to distinct partitions, with constant sample width $w$, and arbitrary split sizes $z' \in [2, z]$, the tree admits $P(B, k)$ total distinct partitions where
$$w^{\frac{k-1}{z-1}} \leq  P(B, k) \leq w^{k - 1} .$$
\end{thm}

The theorem follows from an inductive counting argument (see Appendix~\ref{theorem}). Note that the upper bound is tight only when we sample binary partitions (regardless of balance) and the lower bound is tight when we only sample $z$-partitions and $k = z^n$ for some integer $n$. In practice, we have nodes that do not successfully find $w$ samples and generate duplicate regions. Duplicate leaves decrease the number of distinct district plans, whereas duplicate internal nodes can increase the number because they effectively increase the sample width. 

To successfully achieve the full $w^{k-1}$ distinct plans for $k = 2^n$, one must solve 
$w\sum_{d=0}^{n - 1} (2w)^d$
partition problems involving center selection and the PIP. This is less work than it appears because as $d$ increases, the size of the PIP decreases exponentially. In particular, where the root node has to partition $|B|$ blocks, a node at depth $d$ will only have to partition $|B| / 2^d$ blocks. Again using the binary case, this process will generate $(2w)^{log_2(k)} / 2$ districts which admit $w^{(k-1)}$ unique district plans. This exponential ratio between the number of districts and plans is the targeted column efficiency property.

We call the log of the ratio between the number of generated plans and districts, $\log(\frac{|P|}{|D|})$, the \textit{leverage} of the ensemble. To our knowledge, all existing ensemble approaches actually have negative leverage; that is, they generate more districts than plans. For instance, techniques that generate a fully new plan each iteration (such as greedy agglomerative methods \cite{chen2013unintentional}) have leverage $\ell = \log(\frac{1}{k})$ and Markov chain techniques \cite{chikina2017assessing, deford2019recombination} generate two districts per plan, $\ell = \log(\frac{1}{2}$). Especially for large $k$, having low leverage implies that a technique spends significant computational effort generating districts but that effort does not translate into an efficient exploration of the feasible space.


\subsection{Optimization} \label{optimization}
\paragraph{Objective Function} The second key property of our decoupled design is the flexibility it grants in specifying the objective function. Whereas standard approaches for one-step optimization require the objective be a piecewise-linear function of blocks, our two-step method enables piecewise-linear functions of district-level metrics that can be arbitrary functions of blocks. This property enables our fairness objective. A fair district plan, in our view, is one that, \textit{on average}, calibrates electoral outcomes with the partisan affiliation of a state to minimize the number of \textit{wasted} votes. 


A wasted vote is a vote cast for a losing candidate or a surplus vote cast for the winner over what was needed to win. The efficiency gap (EG) \cite{stephanopoulos2015partisan}, a popular fairness metric, measures the difference between wasted votes for the two parties. The \textit{responsiveness} of a plan is the number $r$ such that a 1\% increase in a party's vote-share is accompanied by a $r\%$ increase in a party's seat-share. Assuming uniform turnout, minimizing the EG coincides with finding a plan with a constant responsiveness of 2 \cite{warrington2018quantifying} (i.e., a 1\% increase in vote-share should always yield a 2\% average increase in seat-share). This is not always possible and scholars debate what the ideal responsiveness should be \cite{mcgann2016gerrymandering}. However, our method is flexible enough to support optimizing any function of votes to seats so we can accommodate alternative definitions of fairness.

To make such probabilistic statements, we require a probabilistic model of partisan affiliation as an input. We use historical precinct returns from statewide elections to estimate the mean and standard deviation of the two-party vote-share for an arbitrary district (see Appendix~\ref{data} for the details of our data sources and preparation). Let $\hat{\nu}$ be the average statewide partisan vote-share, $\nu_i$ be a random variable for the partisan vote-share of district $i$, and $\psi_i$ be a Bernoulli random variable indicating an electoral win or loss for district $i$. 
We assume 
$$\nu_i \sim \mathcal{N}(\mu_i, \sigma_i^2) \quad \quad
\psi_i \sim \mathcal{B}(P(\nu_i > .5))$$ where $\mu_i$ and $\sigma_i^2$ are given by the historical statewide returns of blocks in district $i$. By linearity of expectation, the expected difference between the statewide seat-share and statewide vote-share is the sum of the differences between the expected district-level seat-share and statewide vote-share:
$$
E\left[ \frac{1}{k} \sum_{i=1}^k \hat{\nu} - \psi_i\right]
= \frac{1}{k} \sum_{i=1}^k \hat{\nu} - \left(1 - \Phi\left(\frac{\mu_i - .5}{\sigma_i}\right)\right),
$$
where $\Phi$ is the unit Gaussian cumulative distribution function (CDF). Given the infrequency of elections, in our experiments we use a $t$ random variable CDF, but the model is flexible to support other affiliation models with richer features. Importantly, we can optimize for $h(\hat{\nu}) - \psi_i$ where $h$ gives the ideal mapping from votes to seats enabling optimizing to the shape of arbitrary seat-vote curves. For the efficiency gap, $h(\hat{\nu}) = 2 \hat{\nu} - .5$, assuming uniform turnout. 

Our model is simple by design, making estimation and optimization tractable for millions of arbitrary districts. We do not claim that this is the right model, and in many cases it is the wrong model when incumbency or other specific information is known for a district. However, given the application domain, we believe simplicity is important for adoption, since a more opaque or harder to reason about model would likely be politically more controversial. Regardless, a promising direction for future work is developing more robust affiliation models suitable for use in optimization settings that can generalize to multiple parties or different voting paradigms.


\paragraph{Tree-based Dynamic Programming} When the following conditions hold, the sample tree admits an efficient dynamic programming solution to find the optimal objective.
\begin{enumerate}
    \item[C1.] There are no additional constraints on the composition of districts (e.g., maintaining a specified number of competitive or majority-minority districts) other than those implicitly encoded by the structure of the tree.
    \item[C2.] The objective function is linear (e.g., compactness, competitiveness, but not seat-vote difference)
\end{enumerate}
 The intuition underlying the dynamic program is that for a particular partition of a region, the optimal objective (solution) is the sum (composition) of the optimal objectives (solutions) for each of the subregions in the partition. Written recursively, the optimal objective value is 

$$f^*(R, s) = 
\begin{cases}
\argmin\limits_{i \in w(R)} \sum_{j=1}^{z(i)} f^*(R_{ij}, s_{ij})\ & \text{for} \ \ s > 1 \\
c_{ij} & \text{for} \ \ s = 1
\end{cases}$$

\noindent where $c_{ij}$ is the cost coefficient of the district $R_{ij}$, $z(i)$ is the split size of region partition $i$, and $w(R)$ is the number of partitions sampled.

While this approach is not typically helpful for finding a specific \textit{fair} plan, it is convenient for quickly evaluating the range of the most extreme solutions generated for arbitrary linear metrics. 
This is especially useful in ensemble studies to evaluate how different inputs or constraints effect the range of possible partisan outcomes.

\paragraph{Master Selection Problem} When either condition C1 or C2 above does not hold (most notably when some definition of fairness is the objective), we use the machinery of integer programming to find a family of solutions. This is accomplished by solving a set partitioning master selection problem (MSP) to choose the final set of districts. We solve a separate MSP for each set of leaf nodes that share a parent root partition. In this standard formulation \cite{mehrotra1998optimization}, we have a binary decision variable for every generated district $x_j$ (indexed by $D$) and wish to 

\setcounter{equation}{0}
\begin{align}
    \text{minimize} \quad & \Bigl| \, \sum_{j \in D} c_j x_j \Bigr|  \\
    \text{s.t.}     \quad & \sum_{j \in D} a_{ij} x_j = 1, & \forall i \in B; \\
                    & \sum_{j \in D} x_j = k;  \\
                    & x_j \in \{0, 1\}, & \forall j \in D, 
\end{align}
where $c_j = \mathbf{E}(h(\hat{\nu}) - \psi_j)$ as estimated by our affiliation model and $A$ is the binary block-district matrix with $a_{ij} = 1$ if block $b_i$ is in district $d_j$. Constraint (2.2) enforces that each block is assigned to exactly one district and constraint (2.3) enforces that there are exactly $k$ districts. Constraint (2.3) is strictly necessary only when $k \epsilon_p > 1$, by virtue of the generation process. Any additional linear constraints placed on the composition of districts can be added seamlessly to the formulation (e.g., requiring a minimum number of districts meet a threshold level of competitiveness). Finally, we make the objective linear by introducing and minimizing a new variable $w$, such that
$\sum_{j \in D} c_j x_j \leq \ w$ and $\sum_{j \in D} c_j x_j \geq -w.$

We solve this integer program for the columns generated from each root partition separately for three reasons. By sharding the tree by root partition, the whole pipeline becomes arbitrarily parallelizable. One could set up an arbitrary number of computational workers where each worker runs the SHP generation routine and then solves the MSP for each generated root partition. 
One line of future work is to use optimal transport \cite{abrishami2020geometry} to characterize the convergence behavior of our method to understand when it is unlikely to improve upon the existing solution pool so we can employ more principled terminating conditions.

One concern that might be raised is that by separating subproblems by the root partition, we limit the possible district plans; however, unless an internal node ($R, s$) has an identical copy in another subtree, it is unlikely that two columns generated from different partitions will be compatible. That is, with high probability there does not exist $k-2$ other generated districts that can be included to satisfy constraint (2) of the MSP. One could search for duplicates across the entire tree and augment the column set to include all descendants of all duplicates. However, duplicates are fairly rare and limits the inherent parallelism of our approach.


Finally, in a state where a fair solution exists, there are typically a large number of alternative fair solutions. Since there are many other desirable proprieties of districts plans to trade off, one can filter the range of candidate plans to those that meet a requisite level of fairness, and then consider other properties such as competitiveness, maintaining political boundaries, or not separating communities of interest. Solving one MSP per root partition is a very simple way to get a large number of fair solutions, which are also substantially different from each other. One could use a more sophisticated method \cite{serra2019compact} to find many near optimal solutions for each root partition, but these are more likely to be very similar given the structure imposed by the shared root partition.


\section{Results}
We provide two types of experimental results. The first set of experiments is designed to tune the algorithm's parameters and understand its behavior. We study the performance of different configurations of the SHP generation routine and the impact of varying the PDP parameters. We further provide empirical results of convergence behavior and runtime performance. These results are included in Appendices~\ref{generation_experiment} and \ref{runtime}.

In our main experiment, we generate a large number of districts for every multi-district state to understand the range of possible outcomes using only reasonably shaped districts and compare this to a recombination ensemble baseline. Given the exponential nature of our column space, this constitutes the largest district plan ensemble ever (implicitly) constructed. However, this also makes computing some ensemble-level metrics intractable. Therefore, for some results, we use a pruned sample tree to make enumeration feasible. See Appendix~\ref{ensemble_experiment} for all experiment details. 

\paragraph{Compactness} Compactness has long been used as the objective function in political redistricting research, rationalized by the claim that a compact map drawn with no political input is inherently fair. Although highly noncompact human drawn districts can signal manipulation, there is no \textit{a priori} reason to assume that a randomly drawn compact map will be more fair than a randomly drawn noncompact map. To verify this, we compute the Spearman correlation coefficient \cite{schober2018correlation} for each state between the magnitude of the expected efficiency gap and three different measures of compactness: centralization, Roeck, and cut-edges. Centralization is the average distance a constituent has to walk to the center of their district, averaged over all districts (similar to the dispersion measure of compactness \cite{young1988measuring}); Roeck is the district area divided by the area of the smallest circumscribing circle averaged over all districts; and cut-edges measures the average number of edges that need to be cut in the adjacency graph to form the districts of a plan.

\begin{figure*}
    \includegraphics[width=\linewidth]{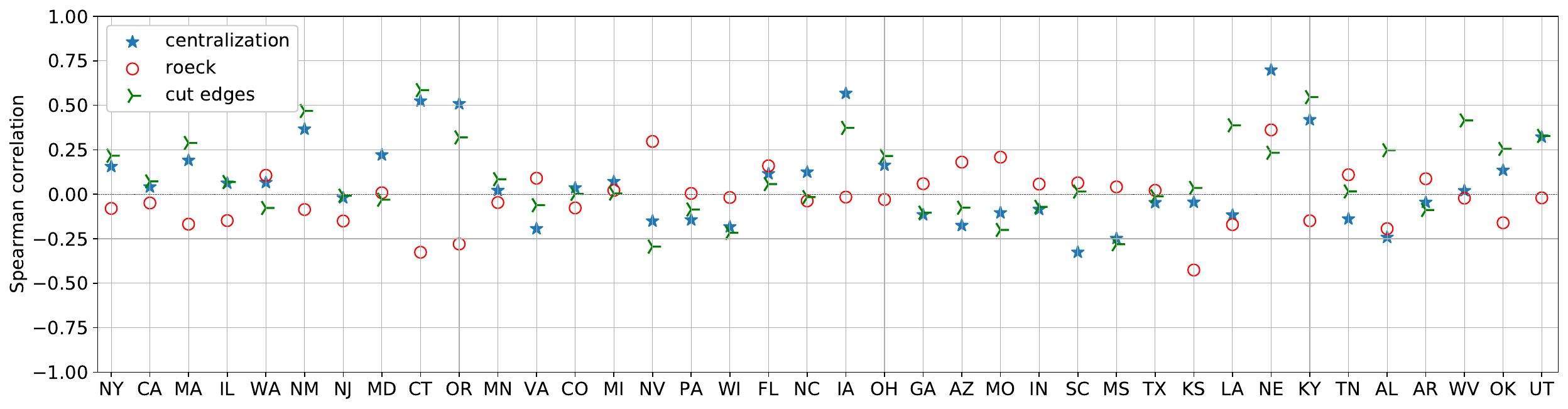}
    \caption{Spearman correlation coefficient between the magnitude of the expected efficiency gap and three different measures of compactness for the 38 states with three or more congressional districts ordered by partisan affiliation.}
    \label{fig:correlation_coeffs}
\end{figure*}

Figure~\ref{fig:correlation_coeffs} and Table~\ref{tab:average_correlation} show that compactness and fairness are orthogonal attributes of a district plan. Any correlation between compactness and fairness is noise stemming from the political geography of a state. For instance in Nebraska, where the correlation is the strongest, the only redistricting decision that affects the expected efficiency gap is whether Omaha is cracked and diluted into two reliably Republican districts (not compact and not fair) or packed into one toss-up district (compact and fair). In many swing states where redistricting decisions can have the most impact, such as Pennsylvania, Wisconsin, and Georgia, the correlation is actually negative. This is because the most compact plans pack the major cities into landslide Democratic districts, while the surrounding suburbs and rural areas have a more efficient placement of Republican voters. Lastly, in the largest states, California and Texas, where no single geographic feature can dominate the correlation, we see that the relationship is almost perfectly independent.

\begin{table}[]
    \centering
\begin{tabular}{lrr}
\toprule
{} &  mean $\rho$ &   $k$-weighted mean $\rho$ \\
\midrule
centralization &     0.074756 &              0.038859 \\
roeck          &    -0.000355 &             -0.008970 \\
cut\_edges      &     0.105347 &              0.066784 \\
\bottomrule
\end{tabular}
    \caption{Average (and $k$-weighted average) Spearman correlation coefficient between the magnitude of expected efficiency gap and three different compactness metrics.}
    \label{tab:average_correlation}
\end{table}

We do not claim that compactness is not important or desirable; compact plans are more likely to represent more coherent communities and make administering elections marginally easier. However, fairness is a far more important consideration and therefore we want to encourage practitioners to treat compactness as a constraint rather than an objective.
\begin{figure*}
    \includegraphics[width=\linewidth]{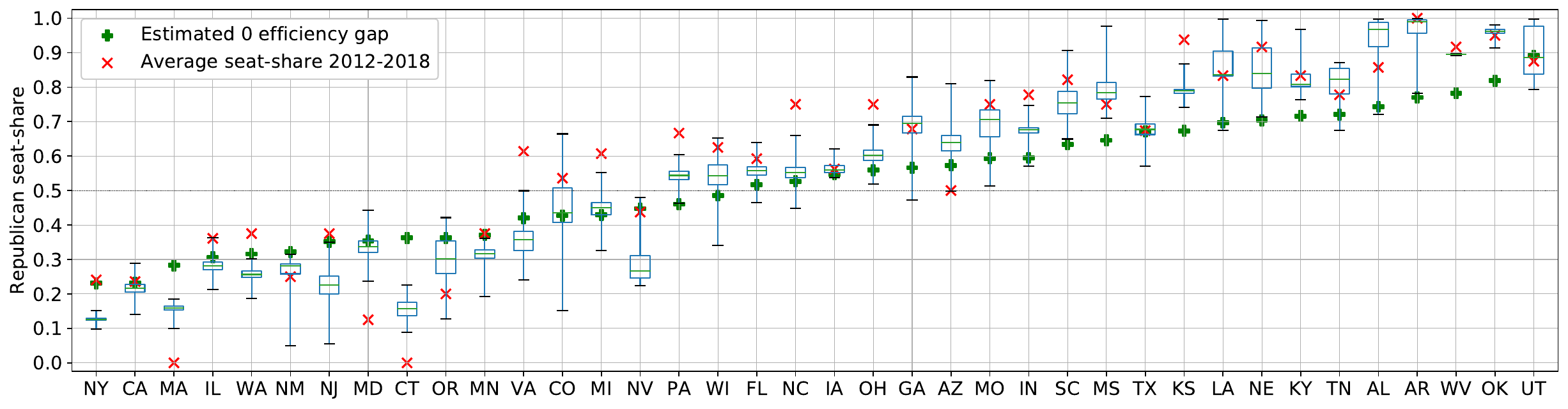}
    \caption{Distribution of expected Republican seat-share of generated plan ensemble for each state with three or more congressional districts ordered by partisan affiliation. The box center indicates the median, with the top and bottom of the box indicating the 75th and 25th percentiles respectively of the down-sampled ensemble. The whiskers denote the minimum and maximum value of any plan found using the tree based dynamic programming optimization algorithm.}
    \label{fig:seat_share_distribution}
\end{figure*}

\begin{figure*}
    \includegraphics[width=0.495\linewidth]{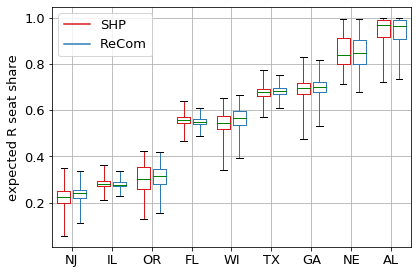}
    \includegraphics[width=0.495\linewidth]{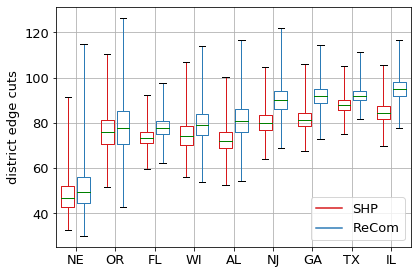}
    \caption{Comparison of expected seat-share (left) and average district edge cuts (right) on ensembles generated with Stochastic Hierarchical Partitioning (SHP) and recombination Markov chains.}
    \label{fig:recom_comparison}
\end{figure*}
\paragraph{Fairness} Unfortunately, fairness, like compactness, does not have a universally agreed-upon characterization; there are different and often competing aspects of fairness, which are difficult to reconcile with the distribution of partisan voters. The political geography of a state, typically defined by a few urban concentrations of more liberal voters punctuating larger and sparser regions of more conservative voters, does not guarantee a fair solution for any reasonable definition of fair. This greatly complicates creating a unified definition of fairness when the number of districts and distribution of voters greatly constraints the space of expected outcomes \cite{duchin2019locating}. We use the efficiency gap \cite{stephanopoulos2015partisan} as our primary metric because it has gained traction as a standard metric in redistricting lawsuits, has an intuitive and compelling interpretation, and is simple to embed in an integer program as we have shown in Section~\ref{optimization}.

\begin{figure}
    \includegraphics[width=\linewidth]{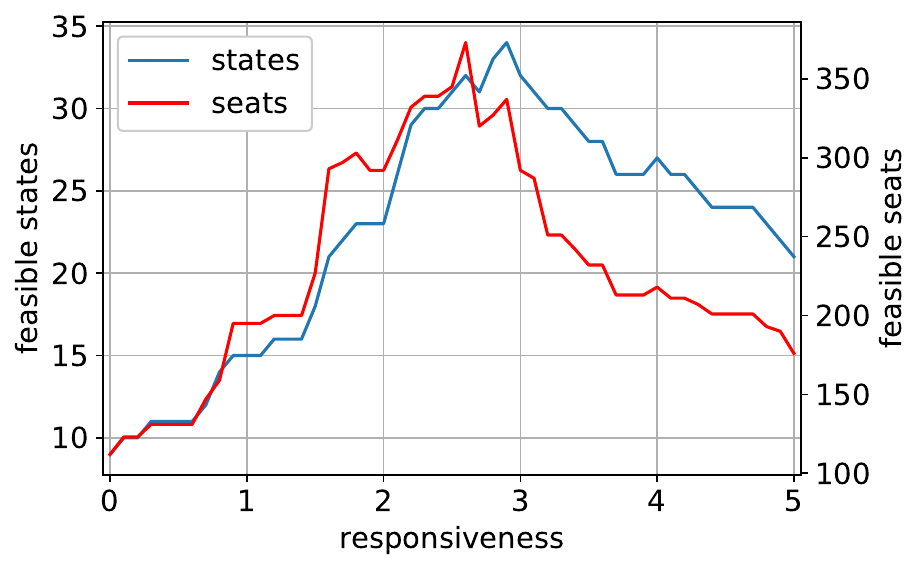}
    \caption{The number of states and total seats that admit a district plan in our ensemble for an expected level of responsiveness.}
    \label{fig:responsivness_feasibility}
\end{figure}


Figure~\ref{fig:seat_share_distribution} shows the range of expected outcomes found by our generation algorithm and includes the point which would minimize the efficiency gap as well as the actual average seat-share over the last redistricting round ordered by partisan affiliation (see Appendix~\ref{ensemble_experiment} for experiment details). By construction, our approach generates only relatively compact districts which we call \textit{natural} districts, so this range can be thought of as the geographic tolerance for natural or unintentional gerrymandering \cite{chen2013unintentional}: reasonably shaped plans that pass the eyeball test but are actually unrepresentative. In every state, we find a plan that is more compact than the current one; in just 7 states the enacted plan is more compact than our median plan, and in 8 states (most of which are well known for gerrymandering) the enacted plan is less compact than any plan we generate (see Appendix~\ref{ensemble_experiment}).

The headline result is that using only natural district boundaries, we can change the expected composition of Democrats in the House of Representatives from 43\% to 62\%. A 20\% swing in the partisan composition of the House is exceptionally consequential, since a body with full control of all congressional boundaries could theoretically create a permanent majority or permanent minority for either party, barring any extreme shifts in partisan affiliation along geographic lines. We run the same experiment with competitiveness (see Appendix~\ref{ensemble_experiment}) and in aggregate achieve plans with natural districts that have between 32 and 106 expected total seat swaps per election.

However, the expected partisan seat range is highly variable. Some states like Colorado admit reasonable district plans that can swing by up to 50\% of total seats, whereas others like West Virginia or Iowa offer very little room to alter the outcome using only natural districts. Consequently, any definition of fairness that relies on an ideal level of responsiveness (e.g., the efficiency gap) will not be feasible in all states without relaxing the compactness or contiguity constraints. Figure~\ref{fig:responsivness_feasibility} shows the number of feasible states (and the corresponding number of seats) that have a solution that achieves a target level of responsiveness. In particular, only 15 states admit a proportional plan where the expected seat-share equals the expected vote-share. Additionally, we found plans with 0 expected efficiency gap for only half of the multi-district states.

\paragraph{Recombination Baseline}
To test our claims about more effective sampling of extreme solutions, we compare ensembles generated with our stochastic hierarchical partitioning (SHP) algorithm and recombination Markov chains \cite{deford2019recombination}, the predominant tool in quantitative redistricting analyses. Recombination chains generate ensembles by iteratively merging two random adjacent districts, sampling a random spanning tree, and choosing a tree cut to balance the populations of the two connected components to create two new districts.

In Figure~\ref{fig:recom_comparison}, we compare the expected seat-share and the edge cuts between ensembles generated with the two methods for a representative sample of states. To make the comparison as meaningful as possible, we generate the same number of distinct districts with each method for each state. In all states but Nebraska, SHP finds a larger range of expected seat-shares (while also being significantly more compact on average). This is expected since small states have very short sample trees and therefore do not benefit as much from more efficient columns as larger states do. Furthermore, by using random spanning trees instead of an inner partition integer program that optimizes for region compactness, recombination chains can sample less compact partitions, which may achieve a more extreme outcome than a natural district would.

\section{Conclusion}
Part methodological exposition, part civic plea, this paper introduces a scalable new method for creating fair political districts. Using the ensembling capability of our algorithm, we performed the largest ever district ensemble study to show the spurious relationship between fairness and compactness. We also illustrated the range of possible outcomes with reasonably shaped districts to better frame the constraints of fairness. We want to emphasize the overall framework more than any specific design decision, as we see the main contribution of this work as offering an alternate paradigm for solving supervised regionalization problems, most notably congressional redistricting, and potentially a more general heuristic for solving other hard computational problems.

Most importantly, we want our work to actually move the needle on redistricting reform. Our core capability is quickly, cheaply, and transparently generating a huge number of legal and fair district plans. We extend an open invitation to collaborate with states in the upcoming redistricting cycle and hope that the maps of the next decade are more representative than those of the last.

\printbibliography

\newpage
\appendix
\pdfoutput=1
\noindent
{\large {\bf Appendix}}

\noindent
All code and data is made publicly available at our \href{https://github.com/fairmandering/fairmandering_paper}{github repository}. All computational experiments were run on a Dell R620 equipped with two Intel Xeon E5-2680 2.70GHz 8-core processors and 96GB of RAM using Gurobi v9.0.2 to solve all integer programs. Throughout, when working with political data, higher values correspond to more Republican voters or a greater advantage to the Republican party, and lower values for Democrats, to maintain the convention that Republicans are on the right and Democrats are on the left.

\begin{figure*}
    \centering
    \includegraphics[width=0.8\linewidth]{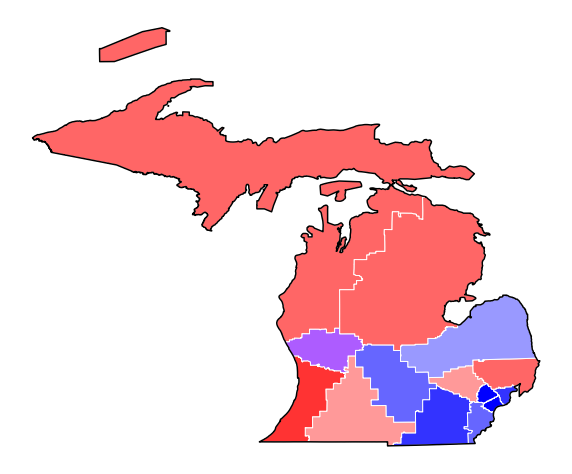}
    \caption{A fair SHP generated map for Michigan.}
    \label{fig:MI}
\end{figure*}

\begin{figure*}
    \centering
    \includegraphics[width=0.9\linewidth]{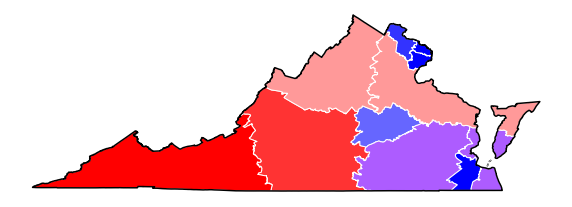}
    \caption{A fair SHP generated map for Virginia.}
    \label{fig:VA}
\end{figure*}

\begin{figure*}
    \centering
    \includegraphics[width=0.65\linewidth]{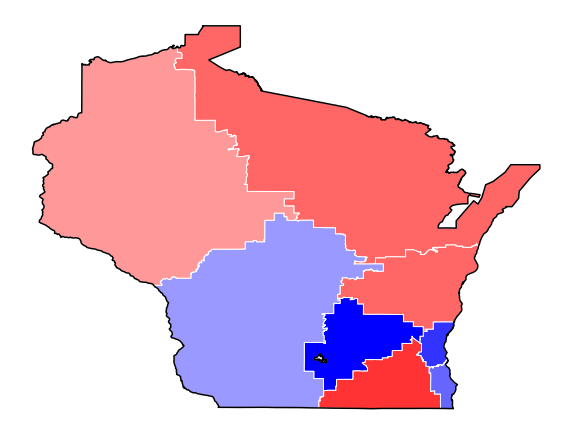}
    \caption{A fair SHP generated map for Wisconsin.}
    \label{fig:WI}
\end{figure*}

\begin{figure*}
    \centering
    \includegraphics[width=0.45\linewidth]{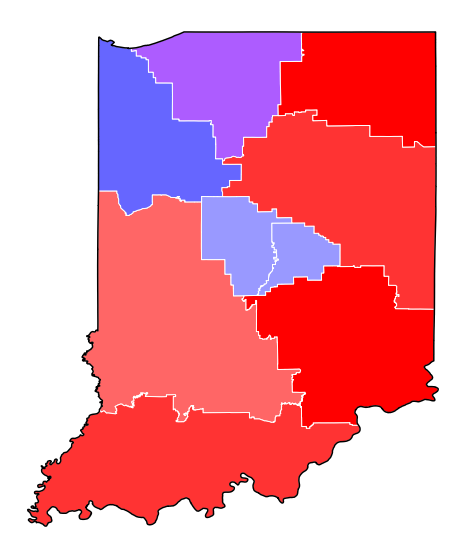}
    \caption{A fair SHP generated map for Indiana.}
    \label{fig:IN}
\end{figure*}

\begin{figure*}
    \centering
    \includegraphics[width=\linewidth]{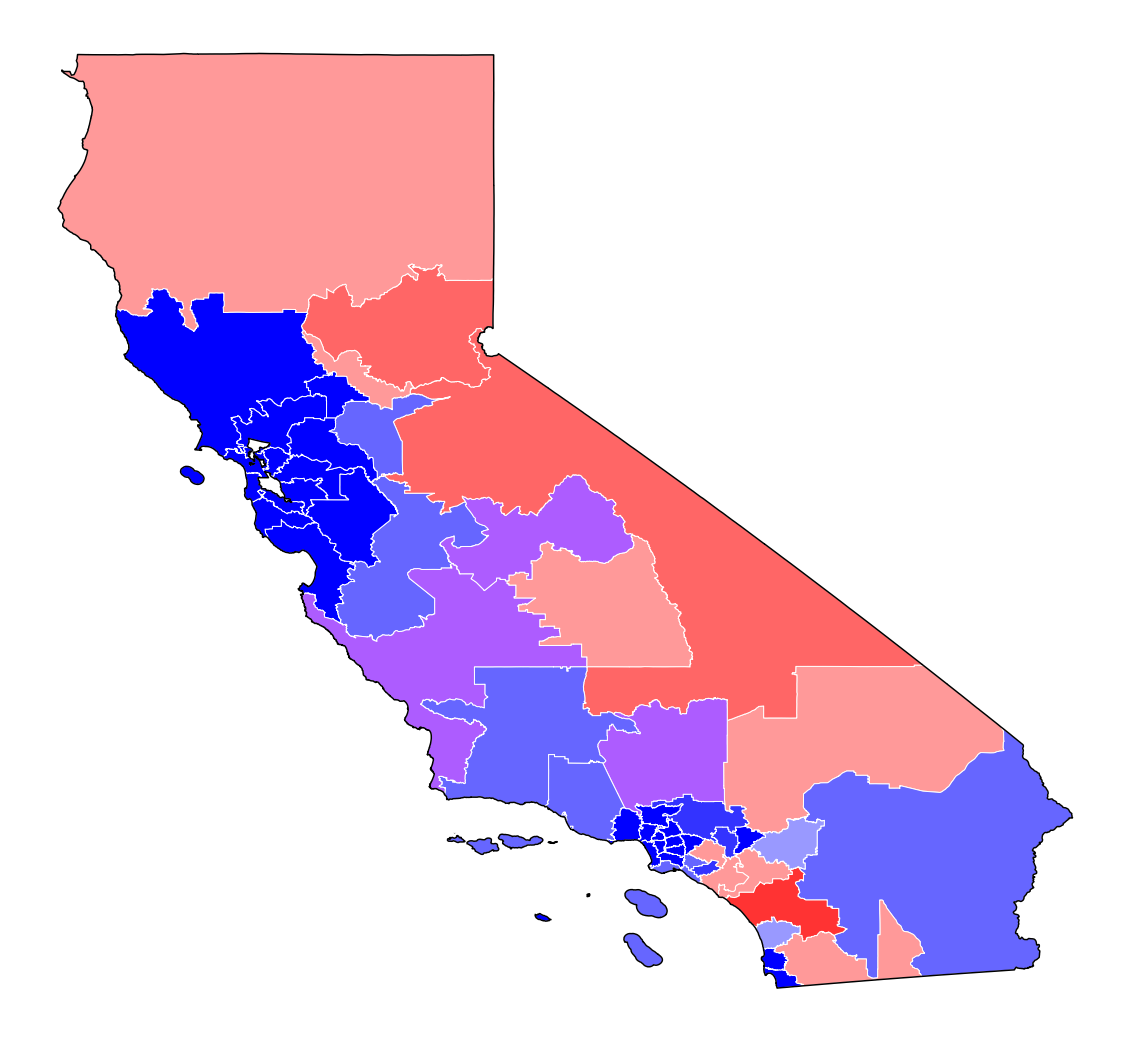}
    \caption{A fair SHP generated map for California.}
    \label{fig:CA}
\end{figure*}

The rest of the appendix is structured as follows. In Section~\ref{data} we describe the data sources and preprocessing steps required to create the data structures necessary for our algorithm. Section~\ref{algo_details} describes a few small additional details regarding the SHP algorithm as well as information regarding the how we randomly sample centers and match these to capacities. Section~\ref{theorem} gives a proof of the theorem in the main text regarding the bounds of the size of the feasible set yielded by a sample tree. Section~\ref{generation_experiment} introduces metrics for measuring ensemble diversity and presents empirical results of the performance of different center selection and capacity matching methods. Additionally, Section~\ref{generation_experiment} shows how the SHP algorithm performs under different PDP parameters such as $\epsilon_p$, the population tolerance, and $k$, the number of districts in a plan. Section~\ref{ensemble_experiment} describes our subsampling algorithm and how we calculate compactness as well as much more detailed qualitative analysis of our results. In Section~\ref{runtime} we give empirical runtime and convergence results of both the partition problems required for generation as well as the master selection problem. Finally, in Figures~\ref{fig:MI}, \ref{fig:VA}, \ref{fig:WI}, \ref{fig:IN}, and \ref{fig:CA} we present a sample of our congressional district plans.

\section{Data} \label{data}
To estimate the partisan affiliation distributions used in our optimization model, we use historical statewide election data gathered from a number of election data repositories  \cite{mgggstates, VOQCHQ_2018, DVN/NH5S2I_2018, DVN/UBKYRU_2019, DVN/UUCWPP_2011, DVN/AWE39N_2011, DVN/KX0YGR_2011, DVN/AN00LH_2011, DVN/WYXFW3_2011}. For each state we take precinct level returns for Presidential, Gubernatorial, Senate, and Attorneys General races, and aggregate these results to census tracts based on overlapping area with precincts. In each of these races, we consider only votes cast for the Democratic and Republican parties, and remove anomalous elections, either those with a strong third-party candidate (e.g., Utah 2016 presidential) or those that are unrepresentative of congressional elections (e.g., Republican Governor of Massachusetts reelected in 2016 with 66\% of the vote in a state that hasn't elected a Republican Representative since 1997). We use state-level returns instead of district-level returns due to the noisiness of congressional elections caused by incumbency advantage and uncontested races. 

\begin{figure*}
    \includegraphics[width=\linewidth]{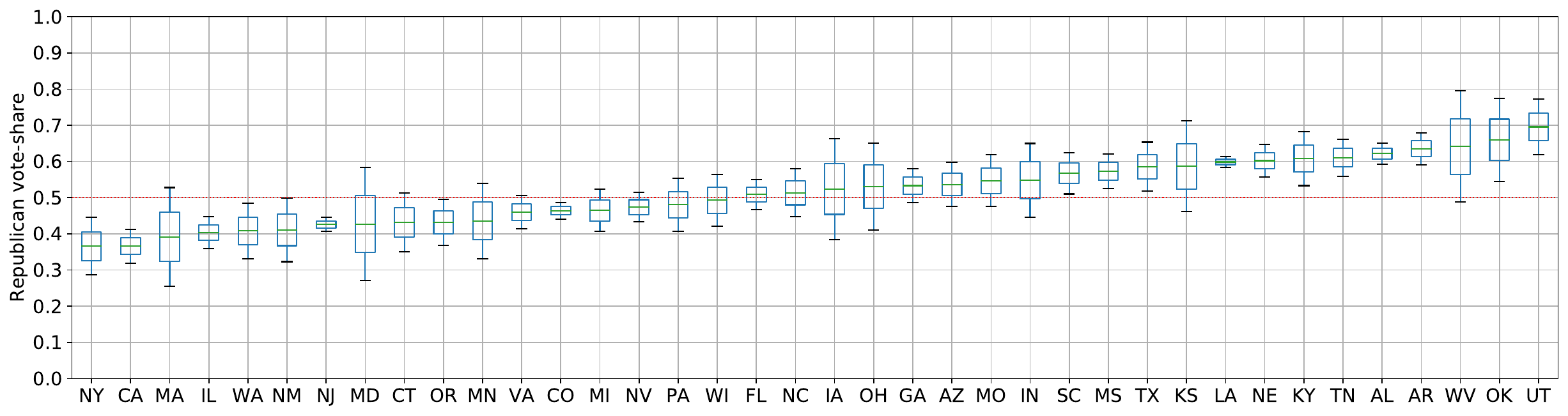}
    \caption{The state-level partisan affiliation for all states with 3 or more districts ordered by mean affiliation. The middle green bar of each box is the sample mean; the box and whisker boundaries measures one and two sample standard deviations respectively.}
    \label{fig:affiliation}
\end{figure*}

For many states, we have only one or two races with complete geography-matched precinct data. To augment our dataset, we combine precinct and county data to infer precinct-level returns for past elections. Specifically, given precinct-level election data for election $Y$ and county-level election data for a different election $X$, we calculate the difference between $Y$ and $X$ at the county-level, and add the difference uniformly to the $Y$ precinct data to infer precinct results for $X$. See Table~\ref{tab:elections} for a list of the elections used in our analysis and Figure~\ref{fig:affiliation} to see the respective state-level vote-share affiliation model.

In addition to election data, our generation algorithm requires a collection of geographic blocks with corresponding population totals and adjacency matrix. For population totals, we use the 2018 American Community Survey 5-year estimates (ACS5). We use the TIGER/Line 2018 census tract shapefiles as our geographic blocks and match this to population and demographic data using geographic entity codes (GEOIDs). With the census shapefiles we compute the block adjacency matrix with PySAL \cite{pysal2007} and iteratively join disconnected regions based on minimum pairwise distance to create one fully connected component. See Table~\ref{tab:state_stats} for adjacency graph and populations statistics.

\begin{table*}
    \centering
    \small
\begin{tabular}{lllp{12cm}}
\toprule
{} & sample mean & sample std &                                                                                                                elections \\
\midrule
AL &       0.622 &      0.014 &                                                                           pres\_2008, senate\_2008, pres\_2012*, pres\_2016* \\
AZ &       0.536 &       0.03 &                                           pres\_2008*, pres\_2012*, senate\_2016, pres\_2016, gov\_2018, senate\_2018, AG\_2018 \\
AR &       0.635 &      0.022 &                                                        pres\_2008*, pres\_2012*, pres\_2016, senate\_2016, gov\_2018, AG\_2018 \\
CA &       0.366 &      0.023 &                                                                                        pres\_2008*, pres\_2012*, pres\_2016 \\
CO &       0.464 &      0.012 &                                                        pres\_2008*, pres\_2012*, senate\_2016, pres\_2016, gov\_2018, AG\_2018 \\
CT &       0.431 &      0.041 &                                                                   pres\_2008*, pres\_2012*, gov\_2018, senate\_2018, AG\_2018 \\
FL &       0.508 &      0.021 &                                                        pres\_2008*, pres\_2012*, pres\_2016, senate\_2016, gov\_2018, AG\_2018 \\
GA &       0.533 &      0.024 &                                                        pres\_2008*, pres\_2012*, pres\_2016, senate\_2016, gov\_2018, AG\_2018 \\
HI &       0.291 &      0.042 &                             pres\_2008*, pres\_2012*, pres\_2016, senate\_2016, senate\_2018, gov\_2018, senate\_2018, gov\_2018 \\
ID &       0.656 &      0.035 &                                                        pres\_2008*, pres\_2012*, pres\_2016, senate\_2016, AG\_2018, gov\_2018 \\
IL &       0.403 &      0.022 &                                                                           pres\_2008*, pres\_2012*, pres\_2016, senate\_2016 \\
IN &       0.547 &      0.051 &                                                                                        pres\_2008, pres\_2012*, pres\_2016* \\
IA &       0.523 &       0.07 &                                                                 pres\_2008*, pres\_2012*, pres\_2016, senate\_2016, gov\_2018 \\
KS &       0.586 &      0.063 &                                                        pres\_2008*, pres\_2012*, pres\_2016, senate\_2016, AG\_2018, gov\_2018 \\
KY &       0.608 &      0.037 &                                                                           pres\_2008*, pres\_2012*, pres\_2016, senate\_2016 \\
LA &       0.598 &      0.007 &                                                                           pres\_2008*, pres\_2012*, pres\_2016, senate\_2016 \\
ME &       0.447 &      0.036 &                                                                              pres\_2008*, pres\_2012*, pres\_2016, gov\_2018 \\
MD &       0.427 &      0.078 &            pres\_2008*, pres\_2012, senate\_2012, gov\_2014, AG\_2014, pres\_2016, senate\_2016, gov\_2018, senate\_2018, AG\_2018 \\
MA &       0.391 &      0.068 &   senate\_2008, pres\_2008, senate\_2010, pres\_2012, senate\_2012, senate\_2013, senate\_2014, pres\_2016, senate\_2018, AG\_2018 \\
MI &       0.465 &      0.029 &                                                        pres\_2008*, pres\_2012*, pres\_2016, senate\_2018, AG\_2018, gov\_2018 \\
MN &       0.436 &      0.052 &            pres\_2008*, pres\_2012, senate\_2012, senate\_2014, gov\_2014, AG\_2014, pres\_2016, senate\_2018, gov\_2018, AG\_2018 \\
MS &       0.573 &      0.024 &                                                            pres\_2008, senate1\_2008, senate2\_2008, pres\_2012*, pres\_2016* \\
MO &       0.547 &      0.036 &                                                                 pres\_2008*, pres\_2012*, pres\_2016, senate\_2016, gov\_2016 \\
NE &       0.602 &      0.023 &                                                                 pres\_2008*, pres\_2012*, pres\_2016, senate\_2018, gov\_2018 \\
NV &       0.474 &       0.02 &                                           pres\_2008*, pres\_2012*, pres\_2016, senate\_2016, senate\_2018, gov\_2018, AG\_2018 \\
NH &       0.495 &      0.029 &                                                       pres\_2008*, pres\_2012*, pres\_2016, senate\_2016, gov\_2016, gov\_2018 \\
NJ &       0.426 &       0.01 &                                                                           pres\_2008, senate\_2008, pres\_2012*, pres\_2016* \\
NM &       0.411 &      0.044 &                                                        pres\_2008*, pres\_2012*, pres\_2016, AG\_2018, gov\_2018, senate\_2018 \\
NY &       0.366 &       0.04 &                                        pres\_2008*, gov\_2010, AG\_2010, senate1\_2010, senate2\_2010, pres\_2012*, pres\_2016* \\
NC &       0.513 &      0.033 &                   senate\_2008, gov\_2008, senate\_2010, pres\_2012, gov\_2012, senate\_2014, gov\_2016, senate\_2016, pres\_2016 \\
OH &        0.53 &       0.06 &                                                                           pres\_2008*, pres\_2012*, pres\_2016, senate\_2016 \\
OK &       0.659 &      0.057 &                                                        pres\_2008*, pres\_2012*, senate\_2016, pres\_2016, gov\_2018, AG\_2018 \\
OR &       0.431 &      0.032 &                                              pres\_2008*, pres\_2012*, gov\_2016, senate\_2016, pres\_2016, AG\_2016, gov\_2018 \\
PA &        0.48 &      0.036 &            pres\_2008*, senate\_2010, gov\_2010, pres\_2012, senate\_2012, AG\_2012, gov\_2014, pres\_2016, senate\_2016, AG\_2016 \\
RI &       0.387 &      0.029 &                                                                 pres\_2008*, pres\_2012*, pres\_2016, senate\_2018, gov\_2018 \\
SC &       0.567 &      0.028 &                                                        pres\_2008*, pres\_2012*, pres\_2016, senate\_2016, gov\_2018, AG\_2018 \\
TN &        0.61 &      0.026 &                                                                                        pres\_2008*, pres\_2012*, pres\_2016 \\
TX &       0.586 &      0.034 &                                                     pres\_2008*, senate\_2012, pres\_2012, gov\_2014, senate\_2014, pres\_2016 \\
UT &       0.696 &      0.038 &                                                               pres\_2008*, pres\_2012*, gov\_2016, senate\_2016, senate\_2018 \\
VA &        0.46 &      0.023 &                                                        pres\_2008*, pres\_2012*, pres\_2016, gov\_2017, AG\_2017, senate\_2018 \\
WA &       0.408 &      0.039 &                                           pres\_2008*, pres\_2012*, pres\_2016, senate\_2016, gov\_2016, AG\_2016, senate\_2018 \\
WV &       0.641 &      0.077 &                                                                                       pres\_2008*, pres\_2012*, pres\_2016* \\
WI &       0.493 &      0.036 &  pres\_2008*, gov\_2012, pres\_2012, senate\_2012, gov\_2014, AG\_2014, pres\_2016, senate\_2016, senate\_2018, gov\_2018, AG\_2018 \\
\bottomrule
\end{tabular}
    \caption{The statewide elections used in our affiliation model and the sample mean and standard deviation of the two-party Republican vote-share. Offices include, president (pres), senate, governor (gov), and attorney general (AG) for years between 2008 and 2018 where granular precinct data is public. Precinct results for elections marked with '*' were inferred using the difference between county level data from an election with precinct level results. In West Virginia, we just used county vote-shares as the estimated tract vote-share for all tracts in a county.}
    \label{tab:elections}
\end{table*}

\begin{table}
    \centering
\begin{tabular}{lrrrr}
\toprule
{} &  districts &  nodes &  edges &  population \\
\midrule
AL &          7 &   1179 &   3273 &     4864680 \\
AZ &          9 &   1526 &   3943 &     6946685 \\
AR &          4 &    686 &   1858 &     2990671 \\
CA &         53 &   8036 &  21935 &    39148760 \\
CO &          7 &   1249 &   3358 &     5531141 \\
CT &          5 &    829 &   2259 &     3581504 \\
FL &         27 &   4199 &  11180 &    20598139 \\
GA &         14 &   1964 &   5478 &    10297484 \\
HI &          2 &    326 &   1450 &     1422029 \\
ID &          2 &    298 &    796 &     1687809 \\
IL &         18 &   3121 &   8262 &    12821497 \\
IN &          9 &   1508 &   4095 &     6637426 \\
IA &          4 &    825 &   2150 &     3132499 \\
KS &          4 &    770 &   2007 &     2908776 \\
KY &          6 &   1115 &   3040 &     4440204 \\
LA &          6 &   1136 &   3124 &     4663616 \\
ME &          2 &    351 &    946 &     1332813 \\
MD &          8 &   1394 &   3691 &     6003435 \\
MA &          9 &   1471 &   4078 &     6830126 \\
MI &         14 &   2767 &   7151 &     9957488 \\
MN &          8 &   1335 &   3632 &     5527358 \\
MS &          4 &    661 &   1768 &     2988762 \\
MO &          8 &   1393 &   3742 &     6090062 \\
NE &          3 &    532 &   1369 &     1904760 \\
NV &          4 &    683 &   1765 &     2922849 \\
NH &          2 &    294 &    781 &     1343622 \\
NJ &         12 &   2004 &   5500 &     8881845 \\
NM &          3 &    499 &   1332 &     2092434 \\
NY &         27 &   4900 &  13288 &    19618453 \\
NC &         13 &   2183 &   6045 &    10155624 \\
OH &         16 &   2946 &   8074 &    11641879 \\
OK &          5 &   1046 &   2764 &     3918137 \\
OR &          5 &    827 &   2267 &     4081943 \\
PA &         18 &   3217 &   8969 &    12791181 \\
RI &          2 &    241 &    631 &     1056611 \\
SC &          7 &   1097 &   3016 &     4955925 \\
TN &          9 &   1497 &   4127 &     6651089 \\
TX &         36 &   5253 &  14048 &    27885195 \\
UT &          4 &    588 &   1608 &     3045350 \\
VA &         11 &   1895 &   5252 &     8413774 \\
WA &         10 &   1446 &   3909 &     7294336 \\
WV &          3 &    484 &   1266 &     1829054 \\
WI &          8 &   1393 &   3745 &     5778394 \\
\bottomrule
\end{tabular}
    \caption{Selected statistics of all multi-district states: the number of districts apportioned in the 2010 redistricting cycle, the population estimated by the 2018 ACS-5, and the number of nodes and edges of the census tract adjacency graph.}
    \label{tab:state_stats}
\end{table}

\section{Algorithm Details} \label{algo_details}
\paragraph{Implementation Details}
There are a number of miscellaneous implementation details that are important for characterizing the SHP algorithm and ensuring that it runs smoothly. First, there are a family of parameters that control the possible shapes of the sample tree: the sample width $w$, minimum split size, the maximum split size, and the maximum sibling capacity disparity. As discussed in the main text, the sample width controls the trade-off between column efficiency and diversity. The max sibling capacity disparity parameter ensures that the tree is relatively balanced by preventing sampling region sizes that are too different. We constrain the size of the maximum sample capacity to be no greater than twice the size of the smallest possible capacity (i.e., 6 and 12 would be the largest allowed split). The min and max split size parameters also help maintain balance by preventing excessively large partitions. We set the min to 2 and max to 5 (or the size of the region if $s<5$) and sample uniformly from this range.

To prevent spending excessive time searching for solutions in regions that do not admit many feasible partitions and, in the worst case, prevent infinite loops, we set a maximum number of sample attempts for each node. In practice, this means picking a constant $maxSamples$ and terminating the sampling loop when either we have attempted $maxSamples$ or we reach the desired number of partitions. If none of the attempts lead to a feasible partition, then we propagate the failure up the tree and prune all related nodes from the sample tree.

Finally, we tighten the population tolerance at the root of the tree and relax it as we descend the tree. Since the optimal solution of an integer program lies at an extreme point, the partitioning IP typically produces regions with populations near the maximum or minimum tolerance. Furthermore, these errors are compounded as we descend the tree, creating potential feasibility issues if the number of blocks is small with respect to the allowed error. To prevent this, we alter the population tolerance constraint depending on the estimated height of the current node. Specifically for a region with size $s$ we set $$\epsilon_p(s) = \epsilon_p / \left \lceil{\log_2(s)}\right \rceil.$$
 
\paragraph{Center Selection}
Center selection methods are designed to sample center blocks that will lead to a diverse set of districts, without compromising feasibility of the partition model or region compactness. We present four different methods for sampling random centers for a region partition: fixed-center, exponential-perturbation, random-iterative and uniform-random.

Fixed-center selection starts by randomly sampling a center uniformly at random from the set of region blocks $c_i \in R$. The rest of the centers are chosen as the output of a weighted $k$-means routine; the input is the list of block centroids with weights equal to the the block population, except for $c_i$ which has weight $w_i = \infty$. For a partition with split size $z$, we use Lloyd's algorithm \cite{lloyd1982least} to approximate the $z$-means of the weighted tract centroids. Finally, we return the $z$ blocks with centroids closest to the $z$ means as the final set of centers.

Exponential-perturbation selection is a generalization of the fixed-center selection method. Rather than randomly perturbing one weight to $\infty$, we perturb each weight with a factor drawn from a heavy-tailed distribution; in this case, we use a Type II Pareto distribution with shape parameter $\alpha_{pareto}$. This has the effect of mostly zeroing out some weights while making others so large that the corresponding centers are effectively fixed. As in the fixed-center selection method, we then run $k$-means weighted by these exponentially perturbed population weights and return the $z$ blocks with centroids closest to the $z$ means as the final set of centers.

Random-iterative selection is a probabilistic method of iteratively sampling well-placed centers. We maintain a set of unassigned blocks $U$, which is initialized to $R$. We start by picking, uniformly at random, a seed block $b_i \in U$. Then randomly choose a center $ c_j \in U$ with probability $\propto$ $d_{ij}^{2}$ and capacity $s$. Order the blocks in $U$ by distance to $c_j$ and remove the $n$ closest blocks such that the sum of their populations is $\approx s \hat{p}$. Finally, set the center to be the new seed and repeat the process until the desired number of centers have been chosen. This is, incidentally, a probabilistic version of the first method proposed for the political redistricting problem \cite{vickrey1961prevention}, but rather than using the assignment as the final partition (as the partition is often disconnected and is rarely compact), we use it to properly space centers in a randomized fashion.

Finally, for uniform-random we sample $z$ centers uniformly at random without replacement from the blocks of a region.
\paragraph{Capacity Matching}
The reader might have noticed that only the random-iterative method makes explicit use of the capacities. Therefore, as a post-processing step for each of these methods, we must assign a capacity to each center. We investigate four ways to do this: using Voronoi or fractional weights, and computing or matching capacities. 

Weights are used to determine the ideal capacity of districts with particular centers. Given an assignment matrix $A$ where $A_{ij}$ gives the assignment weight of block $j$ to center $i$, the center weight vector $w$ is equal to $w(A p) / \hat{p}$.
Voronoi weights are binary and equal to 1 when block $j$ is closest to center $i$ and 0 otherwise. Fractional weights are proportional to the inverse squared distance between centers and blocks. More formally, 
$$A^v_{ij} = \begin{cases} 1 \quad \text{if } i = \operatornamewithlimits{argmin}\limits_{i \in C} d_{ij} \\ 0 \quad \text{else}
\end{cases} 
\quad \quad
A^f_{ij} = \frac{d_{ij}^{-2}}{\sum_{i=1}^z d_{ij}^{-2}}$$
The center weight vector $w$ gives the ideal fractional capacity of a region (i.e., 5.3 districts or, equivalently, center $i$ has ideal region population $ 5.3\hat{p}$). With this vector, we can either match sampled capacities to ideal capacities or round the ideal capacities to integers and discard the sampled capacities. Matching is accomplished by reordering the ideal capacities of $w$ and the sampled capacities $s_1, \dots s_z$ in ascending order and matching center $i$ to $s_i$. Computing capacities requires solving the apportionment problem \cite{balinski2010fair}. We minimize the $L_1$ distance of the center capacities (which must be integers) to the ideal center populations, subject to $\sum_i s_i = s$, and optionally, an additional balance constraint to force more balanced sample trees. {\it A priori}, we prefer matching on the grounds that it uniformly samples capacities (and therefore helps with diversity), but we present experimental results for both methods.

\section{Proof of Theorem} \label{theorem}
\setcounter{thm}{0}
\begin{thm}
Consider a set of blocks $B$ to be partitioned into $k$ districts. For a sample tree with root node $(B, k)$ and with nodes corresponding to distinct partitions, with constant sample width $w$, and arbitrary split sizes $z' \in [2, z]$, the tree admits $P(B, k)$ total distinct partitions where
$$w^{\frac{k-1}{z-1}} \leq  P(B, k) \leq w^{k - 1} .$$
\end{thm}

\begin{@proof}
Consider an arbitrary node in the tree corresponding to a region $R$ to be divided into $s$ districts; we create $w$ distinct partitions of $R$, generating subregions $R_{ij}$ with capacity for $s_{ij}$ districts for $i = 1, \dots, w, \ j = 1, \dots, z(i)$, where $z(i)$ gives the split size for partition $i$ and $\sum_{j=1}^{z(i)} s_{ij} = s$. This yields the recursive formula for the number of distinct partitions of $R$ into $s$ districts
$$ P(R, s) = \begin{cases}
\sum_{i=1}^w \prod_{j=1}^{z(i)} P(R_{ij}, s_{ij})\ & \text{for} \ \ s > 1 \\
1 & \text{for} \ \ s = 1
\end{cases}$$
The product appears because for any region partitioned into subregions, any valid partitioning of the subregions can be composed together to form a valid partitioning of the full region. Therefore, the feasible set of all $s$-partitions of a region rooted at $R$ is the Cartesian product of all feasible $s_{ij}$-partitions of the subregions $R_{ij}, \ j \in \{1, \dots, z(i)\}$, and the size of the set is the product of the sizes of the feasible sets of the subregions. By sampling multiple root partitions of $R$ into $w$ sets of subregions, where the subregions are compatible within the same root partition, but incompatible across different root partitions, the feasible set is the union of the feasible sets (one set per root partition). By the distinctness assumption, the size of the union of the feasible sets is the sum of the sizes of each root partition feasible set. 

To complete the proof we consider an inductive counting argument on the depth of the tree to bound $P(R, s)$. Consider sample partition $i$ of region $(R, s)$ with split width $z(i) \in [2, z]$ where $s > z(i)$ (note that if $s = z(i)$ the number of $s$-partitions is exactly one). By induction, the size of the feasible set for partition $i$ is $\prod_{j=1}^{z(i)} P(R_{ij}, s_{ij})$ where $w^{\frac{s_{ij}-1}{z-1}} \leq P(R_{ij}, s_{ij}) \leq w^{s_{ij} - 1}$. Therefore, the size of the feasible set $\prod_{j=1}^{z(i)} P(R_{ij}, s_{ij})$ is bounded by $$\prod_{j=1}^{z(i)} w^{\frac{s_{ij}-1}{z-1}} \leq \prod_{j=1}^{z(i)} P(R_{ij}, s_{ij}) \leq \prod_{j=1}^{z(i)} w^{s_{ij} - 1}.$$

Since $\sum_{j=1}^{z(i)} s_{ij} = k$ for $i = 1,\dots, w$,
$$\prod_{j=1}^{z(i)} w^{\frac{s_{ij}-1}{z-1}} = w^{\sum_{j=1}^{z(i)} \frac{s_{ij}-1}{z-1}} = w^{\frac{k-z(i)}{z-1}} \quad \text{and}
$$ 
$$ \prod_{j=1}^{z(i)} w^{s_{ij}-1} = w^{\sum_{j=1}^{z(i)} s_{ij}-1} = w^{k-z(i)}.$$
Since $z(i) \in [2, z]$, we can further simplify the bounds to
$$w^{\frac{k-z}{z-1}}  \leq \prod_{j=1}^{z(i)} P(R_{ij}, s_{ij}) \leq w^{k-2}.$$
Finally, because we sample $w$ root partitions, the bound on the size of the full tree is
$$w * w^{\frac{k-z}{z-1}}  \leq \sum_{i=1}^{w} \prod_{j=1}^{z(i)} P(R_{ij}, s_{ij}) \leq w * w^{k-2}$$
which simplifies to
$$w^{\frac{k-1}{z-1}} \leq P(R, s) \leq w^{k-1}$$
as desired. 
\end{@proof}

\section{Generation Experiments} \label{generation_experiment}

We test different variations of our generation algorithm using Illinois and North Carolina. We use these two states because they have contrasting political affiliations and geographies. Illinois is traditionally a blue state dominated by one huge city on the geographic exterior while North Carolina is traditionally a red state with many smaller urban areas in the interior.

\paragraph{Varying SHP parameters}
Our first set of experiments test the performance of different center selection and capacity assignment methods to understand the trade-offs between compactness, feasibility, and diversity. For every combination of capacity assignment methods and center selection methods we attempt to sample the root node 100 times and each internal node 3 times. For North Carolina and Illinois respectively, we set the number of district to be 13 and 18, their 2010 apportionment of congressional seats.

Consider Tables~\ref{tab:generation-stats}, \ref{tab:averaged-generation-stats}, and \ref{tab:pop_tol}; the columns from left to right are as follows: $\ell$ is the exponential leverage or $\log_{10}(|F|/|D|)$ which measures the efficiency of the column space. \%infeas is the percent of partition problems that did not have a feasible solution, requiring the resampling of centers and then resolving the partition IP. A measure of compactness, $\mu_{walk}$ is the average district centralization: the average distance in kilometers that a constituent would need to walk to the center of the district averaged across all generated districts. \%dup gives the percentage of duplicate districts generated. 

The next 5 columns are all district diversity measures based on manipulations of the block-district matrix $A$. Since $A$ is almost the entire constraint matrix of the master selection problem, it is a natural and important object to analyze. $H(b_i|b_j)$ is the average binary conditional entropy of the probability that block $i$ is in a district, given block $j$ is in the district. Average district similarity (ADS) is the average pairwise Jaccard similarity (intersection over union) of all districts (multiplied by $k$ to help cross-state comparison). Using the singular value decomposition of $A$, $\sigma_{x}$ is equal to the ratio between the rank needed to produce an $x\%$ approximation and the original rank of $A$. $\lambda_2$ is the second eigenvalue of the normalized Laplacian of the bipartite graph of blocks with edge weights equal to the probability they co-occur together. The second eigenvalue of the normalized graph Laplacian is well known to approximate the minimum graph cut \cite{barnes1982algorithm}. These metrics are all different ways of measuring the regularity of districts by analyzing the block-level cooccurence behavior.

Finally, MCD (maximum compactness disparity) and ESR (expected seat range) are based on the dynamic programming minimum and maximum solutions found. The MCD is quotient of the number of cut-edges of the most and least compact plan of all plans found in the enumeration. The ESR is the maximum range of seat advantage found by the enumeration, using the voter affiliation model. Higher values for $\ell$, $H(b_i|b_j)$, $\sigma_x$, $\lambda_2$, MCD, and ESR are better (the first indicating efficiency, and the latter five indicating diversity) whereas lower values of \%infeas, $\mu_{walk}$ and \%dup are better (indicating well-spaced centers, compactness, and diversity respectively).

Table~\ref{tab:generation-stats} shows the raw totals for each combination of center selection and capacity matching method. Table~\ref{tab:averaged-generation-stats} averages Table~\ref{tab:generation-stats} across each center selection and capacity matching method to more easily compare methods and isolate performance differences.

There is a continuum across the center selection methods regarding how constrained the placement of centers is, both in absolute terms and with respect to the placement of other centers. At one extreme, there is uniform random, which is completely unconstrained. As expected, this leads to the most diverse districts (highest $H(b_i|b_j), \ \sigma_{50},\ \sigma_{99},\ \lambda_2$, MCD, ESR and lowest ADS) but by far the worst average compactness and overall partition feasibility. At the other extreme, there is fixed center and $1_{pareto}$-perturbation, which generally have lower diversity, but are more compact and almost never generate infeasible centers. By using $k$-means as a subroutine, it naturally forces the centers further apart, even with one or more fixed centers and extreme perturbation. It also constrains where centers are placed, since $k$-means will never place a center on the extreme boundary of a state. In between these extremes is random-iterative selection, which uses a weighted random assignment mechanism to ensure that centers aren't placed too closely together, but yields more randomness than $k$-means-based approaches. By combining $k$-means randomization techniques, fixed-center and Pareto perturbation (FC+1-P) also scores well across all metrics achieving a balance between diversity, feasibility, and compactness. In the remaining experiments, we use the FC+1-P technique to generate centers.

Computing versus matching center capacities also makes a significant difference across metrics. When we match capacities, it forces the tree to be more balanced with fewer singleton nodes higher up the tree, yielding higher overall leverage. Matching also leads to better diversity across all our diversity metrics. The only upside for computing capacities is that it leads to fewer infeasible internal nodes and fewer duplicate leaf nodes. Regarding the capacity weights, there is little difference between fractional and Voronoi weights. In Illinois, it appears fractional weights perform slightly better, but in North Carolina, Voronoi weights perform slightly better. For the remaining experiments, we use matched capacities with Voronoi weights (V+M).

\begin{table*}
\small
\vspace{-1em}
\centering
\begin{tabular}{lllrrrrrrrrrrr}
\toprule
   &    &     &  $\ell$ &  \%infeas &  \%dup &  $H(b_i|b_j)$ &    ADS &  $\sigma_{50}$ &  $\sigma_{99}$ &  $\lambda_2$ &  $\mu_{walk}$ &    MCD &    ESR \\
State & Centers & Caps &         &           &        &               &        &                &                &              &               &        &        \\
\midrule
IL & 0.5-P & F+C &    2.40 &      0.23 &   0.31 &        0.0917 &  1.050 &          0.104 &          0.836 &        0.026 &         11.32 &  1.530 &  1.793 \\
   &    & F+M &    2.77 &      0.75 &   0.82 &        0.0964 &  0.698 &          0.105 &          0.893 &        0.027 &         23.25 &  1.574 &  2.060 \\
   &    & V+C &    2.41 &      0.30 &   0.29 &        0.0915 &  1.046 &          0.103 &          0.833 &        0.024 &         11.39 &  1.513 &  2.044 \\
   &    & V+M &    2.84 &      0.61 &   0.90 &        0.0964 &  0.700 &          0.104 &          0.890 &        0.026 &         22.86 &  1.502 &  2.064 \\
   & 1-P & F+C &    2.46 &      0.01 &   0.67 &        0.0826 &  1.120 &          0.096 &          0.794 &        0.019 &         10.51 &  1.459 &  1.667 \\
   &    & F+M &    2.78 &      0.38 &   1.86 &        0.0912 &  0.733 &          0.099 &          0.874 &        0.024 &         22.00 &  1.582 &  2.071 \\
   &    & V+C &    2.43 &      0.01 &   0.66 &        0.0837 &  1.145 &          0.096 &          0.799 &        0.021 &         10.40 &  1.406 &  1.755 \\
   &    & V+M &    2.75 &      0.22 &   1.79 &        0.0911 &  0.745 &          0.098 &          0.868 &        0.023 &         21.67 &  1.587 &  1.886 \\
   & FC & F+C &    2.44 &      0.01 &   0.76 &        0.0822 &  1.114 &          0.094 &          0.786 &        0.018 &         10.71 &  1.408 &  1.680 \\
   &    & F+M &    2.78 &      0.07 &   1.89 &        0.0909 &  0.739 &          0.100 &          0.868 &        0.023 &         21.43 &  1.490 &  2.224 \\
   &    & V+C &    2.43 &      0.00 &   0.71 &        0.0823 &  1.121 &          0.095 &          0.789 &        0.019 &         10.73 &  1.418 &  1.958 \\
   &    & V+M &    2.85 &      0.08 &   1.75 &        0.0913 &  0.737 &          0.099 &          0.872 &        0.023 &         21.63 &  1.555 &  1.957 \\
   & FC+1-P & F+C &    2.41 &      0.07 &   0.42 &        0.0867 &  1.097 &          0.099 &          0.809 &        0.022 &         10.76 &  1.425 &  2.142 \\
   &    & F+M &    2.83 &      0.19 &   1.25 &        0.0933 &  0.715 &          0.102 &          0.880 &        0.025 &         22.88 &  1.606 &  2.071 \\
   &    & V+C &    2.37 &      0.07 &   0.31 &        0.0867 &  1.087 &          0.099 &          0.811 &        0.021 &         10.92 &  1.443 &  1.833 \\
   &    & V+M &    2.78 &      0.32 &   1.10 &        0.0935 &  0.709 &          0.102 &          0.881 &        0.026 &         22.18 &  1.522 &  2.162 \\
   & RI & F+C &    2.18 &      0.18 &   0.42 &        0.0919 &  1.030 &          0.104 &          0.839 &        0.026 &         11.55 &  1.512 &  1.939 \\
   &    & F+M &    2.81 &      0.31 &   1.01 &        0.0943 &  0.706 &          0.104 &          0.889 &        0.024 &         22.46 &  1.489 &  2.502 \\
   &    & V+C &    2.36 &      0.29 &   0.48 &        0.0921 &  1.039 &          0.104 &          0.839 &        0.026 &         11.60 &  1.492 &  2.017 \\
   &    & V+M &    2.85 &      0.42 &   1.04 &        0.0945 &  0.701 &          0.105 &          0.891 &        0.024 &         22.63 &  1.606 &  2.141 \\
   & UR & F+C &    2.52 &      5.60 &   0.33 &        0.1028 &  0.731 &          0.116 &          0.900 &        0.041 &         20.02 &  1.687 &  2.165 \\
   &    & F+M &    2.85 &      6.48 &   0.67 &        0.1047 &  0.651 &          0.112 &          0.912 &        0.038 &         27.04 &  1.645 &  2.375 \\
   &    & V+C &    2.51 &      5.30 &   0.41 &        0.1021 &  0.734 &          0.115 &          0.896 &        0.039 &         19.36 &  1.643 &  2.094 \\
   &    & V+M &    2.84 &      7.57 &   0.87 &        0.1051 &  0.652 &          0.112 &          0.911 &        0.038 &         26.45 &  1.778 &  2.183 \\
   \midrule
NC & 0.5-P & F+C &    1.27 &      0.10 &   1.19 &        0.1127 &  0.802 &          0.100 &          0.843 &        0.029 &         29.05 &  1.658 &  2.117 \\
   &    & F+M &    1.51 &      0.31 &   2.07 &        0.1158 &  0.734 &          0.099 &          0.855 &        0.033 &         31.66 &  1.674 &  2.237 \\
   &    & V+C &    1.33 &      0.26 &   1.09 &        0.1111 &  0.811 &          0.099 &          0.840 &        0.028 &         28.69 &  1.540 &  2.143 \\
   &    & V+M &    1.47 &      0.16 &   2.05 &        0.1149 &  0.729 &          0.099 &          0.858 &        0.031 &         31.95 &  1.528 &  2.148 \\
   & 1-P & F+C &    1.27 &      0.02 &   1.86 &        0.1011 &  0.883 &          0.090 &          0.805 &        0.022 &         27.10 &  1.504 &  1.855 \\
   &    & F+M &    1.49 &      0.02 &   3.26 &        0.1077 &  0.766 &          0.092 &          0.827 &        0.028 &         30.50 &  1.603 &  2.098 \\
   &    & V+C &    1.19 &      0.00 &   1.43 &        0.1009 &  0.873 &          0.090 &          0.804 &        0.020 &         27.34 &  1.523 &  1.892 \\
   &    & V+M &    1.51 &      0.02 &   3.24 &        0.1062 &  0.773 &          0.093 &          0.831 &        0.028 &         30.18 &  1.536 &  1.923 \\
   & FC & F+C &    1.28 &      0.00 &   1.82 &        0.1013 &  0.878 &          0.090 &          0.810 &        0.021 &         27.27 &  1.543 &  1.907 \\
   &    & F+M &    1.50 &      0.00 &   3.12 &        0.1064 &  0.771 &          0.091 &          0.827 &        0.028 &         30.21 &  1.585 &  1.986 \\
   &    & V+C &    1.22 &      0.00 &   1.41 &        0.1012 &  0.879 &          0.088 &          0.809 &        0.020 &         27.27 &  1.587 &  1.852 \\
   &    & V+M &    1.52 &      0.02 &   3.38 &        0.1057 &  0.770 &          0.090 &          0.818 &        0.027 &         30.33 &  1.544 &  1.964 \\
   & FC+1-P & F+C &    1.32 &      0.07 &   1.15 &        0.1062 &  0.851 &          0.095 &          0.829 &        0.024 &         27.80 &  1.611 &  1.872 \\
   &    & F+M &    1.48 &      0.13 &   2.52 &        0.1107 &  0.748 &          0.095 &          0.843 &        0.028 &         31.15 &  1.628 &  2.372 \\
   &    & V+C &    1.25 &      0.13 &   1.15 &        0.1050 &  0.847 &          0.094 &          0.821 &        0.023 &         27.81 &  1.530 &  1.973 \\
   &    & V+M &    1.49 &      0.06 &   2.41 &        0.1115 &  0.744 &          0.096 &          0.846 &        0.030 &         31.28 &  1.642 &  2.069 \\
   & RI & F+C &    1.22 &      0.29 &   1.20 &        0.1138 &  0.802 &          0.101 &          0.842 &        0.034 &         28.74 &  1.624 &  2.146 \\
   &    & F+M &    1.58 &      0.24 &   2.11 &        0.1178 &  0.723 &          0.099 &          0.861 &        0.037 &         32.33 &  1.607 &  2.231 \\
   &    & V+C &    1.26 &      0.28 &   1.08 &        0.1137 &  0.819 &          0.101 &          0.838 &        0.033 &         28.48 &  1.525 &  2.228 \\
   &    & V+M &    1.50 &      0.34 &   2.05 &        0.1168 &  0.731 &          0.098 &          0.857 &        0.035 &         31.68 &  1.585 &  2.330 \\
   & UR & F+C &    1.25 &      4.19 &   0.89 &        0.1313 &  0.710 &          0.112 &          0.893 &        0.043 &         33.67 &  1.798 &  2.572 \\
   &    & F+M &    1.49 &      5.75 &   1.04 &        0.1356 &  0.671 &          0.108 &          0.901 &        0.049 &         37.01 &  1.827 &  2.329 \\
   &    & V+C &    1.24 &      4.70 &   0.69 &        0.1311 &  0.719 &          0.112 &          0.889 &        0.046 &         33.29 &  1.874 &  2.283 \\
   &    & V+M &    1.49 &      5.34 &   0.89 &        0.1346 &  0.673 &          0.109 &          0.901 &        0.047 &         36.92 &  1.772 &  2.230 \\
\bottomrule
\end{tabular}
\caption{Generation summary statistics for Illinois and North Carolina: leverage ($\ell$), percent infeasible of attempted partition problems (\%infeas), percent of districts that are duplicates (\%dup), the conditional binary entropy of two blocks appearing in the same district ($H(b_i|b_j)$), the average district similarity (ADS), the rank ratio of the block district matrix singular value need to reach the 50\% and 99\% reconstruction accuracy ($\sigma_{50}, \ \sigma_{99}$), the second eigenvalue of the normalized graph Laplacian with weights given by the block district matrix ($\lambda_2$), the average distance required to walk to the center of a district ($\mu_{walk}$), the maximum compactness disparity (MCD), and the expected seat range (ESR). Capacity methods: F=fractional, V=Voronoi, C=compute, M=match. Center selection methods: $x$-P=$\alpha_{pareto}$-perturbation, FC=fixed-center, RI=random-iterative, UR=uniform-random.}
\label{tab:generation-stats}
\end{table*}

\begin{table*}
\sisetup{round-mode=places,round-precision=3, table-format=1.3}
\small
\centering
\begin{tabular}{llSSSSSSSSSSSS}
\toprule
   &     &  {{$\ell$}} &  {{\%infeas}} &   {{\%dup}} &  {{$H(b_i|b_j)$}} &     {{ADS}} &  {{$\sigma_{50}$}} &  {{$\sigma_{99}$}} &  {{$\lambda_2$}} &  {{$\mu_{walk}$}} &     {{MCD}} &     {{ESR}} \\
State & Method &         &           &         &               &         &                &                &              &               &         &         \\
\midrule
IL & 0.5-P &  2.6050 &    0.4725 &  0.5800 &        0.0940 &  0.8735 &         0.1040 &         0.8630 &       0.0258 &       17.2050 &  1.5298 &  1.9902 \\
   & 1-P &  2.6050 &    0.1550 &  1.2450 &        0.0872 &  0.9358 &         0.0972 &         0.8338 &       0.0217 &       16.1450 &  1.5085 &  1.8448 \\
   & FC &  2.6250 &    0.0400 &  1.2775 &        0.0867 &  0.9278 &         0.0970 &         0.8288 &       0.0207 &       16.1250 &  1.4677 &  1.9548 \\
   & FC+1-P &  2.5975 &    0.1625 &  0.7700 &        0.0900 &  0.9020 &         0.1005 &         0.8452 &       0.0235 &       16.6850 &  1.4990 &  2.0520 \\
   & RI &  2.5500 &    0.3000 &  0.7375 &        0.0932 &  0.8690 &         0.1042 &         0.8645 &       0.0250 &       17.0600 &  1.5248 &  2.1498 \\
   & UR &  2.6800 &    6.2375 &  0.5700 &        0.1037 &  0.6920 &         0.1138 &         0.9048 &       0.0390 &       23.2175 &  1.6882 &  2.2042 \\
NC & 0.5-P &  1.3950 &    0.2075 &  1.6000 &        0.1136 &  0.7690 &         0.0992 &         0.8490 &       0.0302 &       30.3375 &  1.6000 &  2.1612 \\
   & 1-P &  1.3650 &    0.0150 &  2.4475 &        0.1040 &  0.8238 &         0.0912 &         0.8168 &       0.0245 &       28.7800 &  1.5415 &  1.9420 \\
   & FC &  1.3800 &    0.0050 &  2.4325 &        0.1036 &  0.8245 &         0.0898 &         0.8160 &       0.0240 &       28.7700 &  1.5648 &  1.9272 \\
   & FC+1-P &  1.3850 &    0.0975 &  1.8075 &        0.1084 &  0.7975 &         0.0950 &         0.8348 &       0.0262 &       29.5100 &  1.6027 &  2.0715 \\
   & RI &  1.3900 &    0.2875 &  1.6100 &        0.1155 &  0.7687 &         0.0998 &         0.8495 &       0.0348 &       30.3075 &  1.5852 &  2.2338 \\
   & UR &  1.3675 &    4.9950 &  0.8775 &        0.1332 &  0.6932 &         0.1102 &         0.8960 &       0.0462 &       35.2225 &  1.8178 &  2.3535 \\
  \midrule
IL & F+C &  2.4017 &    1.0167 &  0.4850 &        0.0897 &  1.0237 &         0.1022 &         0.8273 &       0.0253 &       12.4783 &  1.5035 &  1.8977 \\
   & F+M &  2.8033 &    1.3633 &  1.2500 &        0.0951 &  0.7070 &         0.1037 &         0.8860 &       0.0268 &       23.1767 &  1.5643 &  2.2172 \\
   & V+C &  2.4183 &    0.9950 &  0.4767 &        0.0897 &  1.0287 &         0.1020 &         0.8278 &       0.0250 &       12.4000 &  1.4858 &  1.9502 \\
   & V+M &  2.8183 &    1.5367 &  1.2417 &        0.0953 &  0.7073 &         0.1033 &         0.8855 &       0.0267 &       22.9033 &  1.5917 &  2.0655 \\
NC & F+C &  1.2683 &    0.7783 &  1.3517 &        0.1111 &  0.8210 &         0.0980 &         0.8370 &       0.0288 &       28.9383 &  1.6230 &  2.0782 \\
   & F+M &  1.5083 &    1.0750 &  2.3533 &        0.1157 &  0.7355 &         0.0973 &         0.8523 &       0.0338 &       32.1433 &  1.6540 &  2.2088 \\
   & V+C &  1.2483 &    0.8950 &  1.1417 &        0.1105 &  0.8247 &         0.0973 &         0.8335 &       0.0283 &       28.8133 &  1.5965 &  2.0618 \\
   & V+M &  1.4967 &    0.9900 &  2.3367 &        0.1150 &  0.7367 &         0.0975 &         0.8518 &       0.0330 &       32.0567 &  1.6012 &  2.1107 \\
\bottomrule
\end{tabular}
\caption{Generation summary statistics averaged across center selection method (top) and capacity matching method (bottom) for Illinois and North Carolina: leverage ($\ell$), percent infeasible of attempted partition problems (\%infeas), percent of districts that are duplicates (\%dup), the conditional binary entropy of two blocks appearing in the same district ($H(b_i|b_j)$), the average district similarity (ADS), the rank ratio of the block district matrix singular value need to reach the 50\% and 99\% reconstruction accuracy ($\sigma_{50}, \ \sigma_{99}$), the second eigenvalue of the normalized graph Laplacian with weights given by the block district matrix ($\lambda_2$), the average distance required to walk to the center of a district ($\mu_{walk}$), the maximum compactness disparity (MCD), and the expected seat range (ESR). Capacity methods: F=fractional, V=Voronoi, C=compute, M=match. Center selection methods: $x$-P=$\alpha_{pareto}$-perturbation, FC=fixed-center, RI=random-iterative, UR=uniform-random.}
\label{tab:averaged-generation-stats}
\end{table*}

\begin{figure*}
    \centering
      \includegraphics[width=0.495\linewidth]{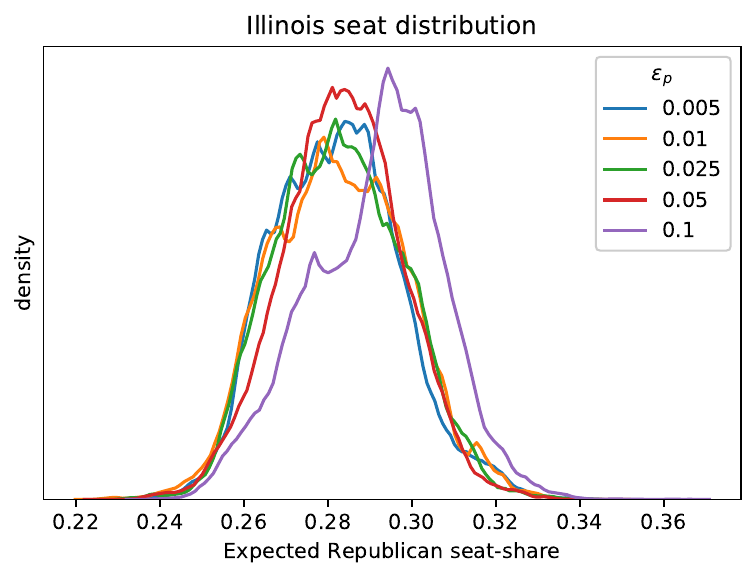}
      \includegraphics[width=0.495\linewidth]{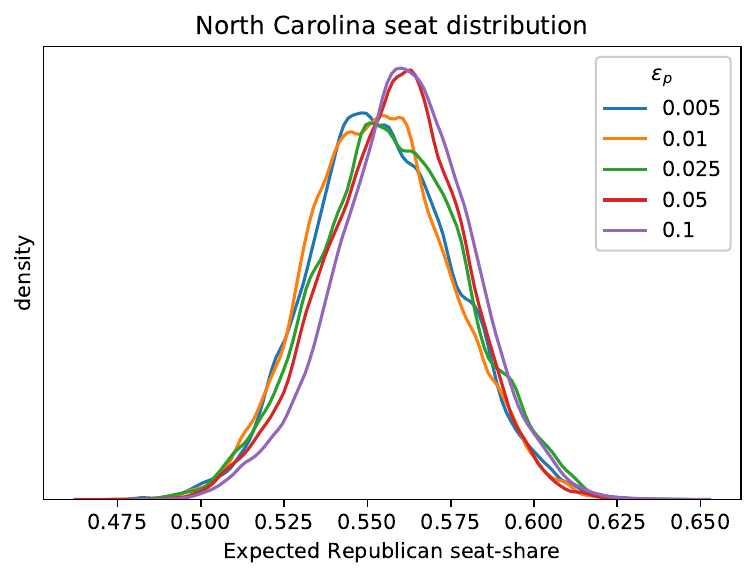}
    \caption{Empirical distribution of seat advantage with varying population tolerance $\epsilon_p$.}
    \label{fig:seat_advantage_pop_tol}
\end{figure*}

\paragraph{Varying PDP parameters} The next experiment is designed to analyze the effect of varying the population tolerance $\epsilon_p$; it is set to .01 in all other experiments. For the reported trials in Table~\ref{tab:pop_tol}, we sample the root node 200 times and each child node 3 times using the Voronoi-match .5-perturbation SHP configuration. We were pleasantly surprised by the fact that there is no substantial difference in the feasibility, duplicate percentage, or diversity of districts generated with $\epsilon_p = .005$ or $\epsilon_p = .1$. Additionally, as depicted in Figure~\ref{fig:seat_advantage_pop_tol}, changing the population tolerance barely affects the resulting seat advantage distribution. The only noticeable impact is with a very high population tolerance, the ensemble favors the Republican party slightly. We believe this happens because our algorithm optimizes with the dispersion measure of compactness, which rewards higher population urban districts more than higher population rural districts. Consequently, the districts we generate will have higher population for more urban (and generally more liberal) districts and lower population for more rural (and more conservative) districts, leading to a less efficient allocation of Democratic voters.

Our last generation experiment is to vary $k$, the number of seats in a plan. Table~\ref{tab:vary_k_generation} shows the parameters of generation in addition to duplicate, feasibility, and leverage statistics. Our empirical leverage is always slightly below the average of the theoretical lower and upper bounds. This is as expected because on average we sample a split size slightly below the average of $[2, z_{max}]$ because the partition size is bounded by the node capacity. The plans/second column shows just how powerful our method is when we scale to large values of $k$. It is a rare moment in algorithms research when exponents work for you and not against you.

\begin{figure*}
    \centering
      \includegraphics[width=0.495\linewidth]{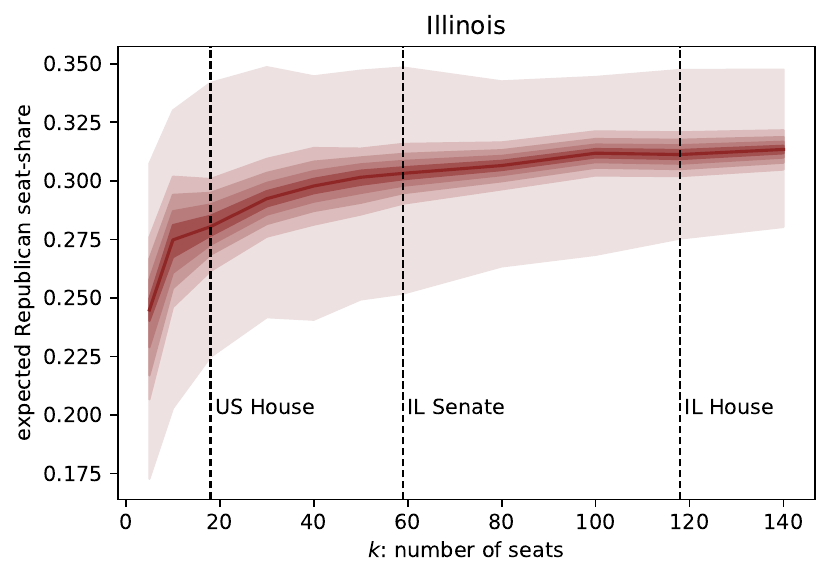}
      \includegraphics[width=0.495\linewidth]{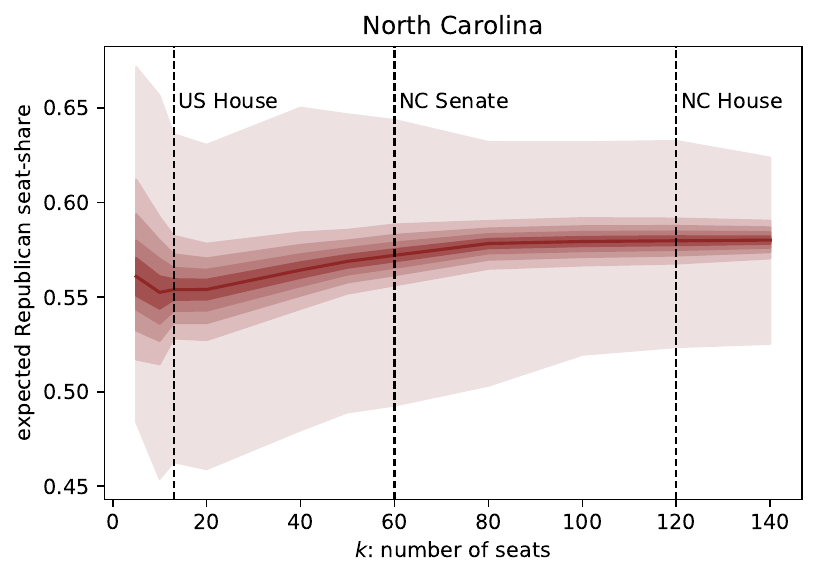}
    \caption{The ensemble distribution of seat-share outcomes for different values of $k$ for Illinois and North Carolina.}
    \label{fig:vary_k}
\end{figure*}

Figure~\ref{fig:vary_k} shows the resulting distribution where color bands encode distribution deciles. This is consistent with other work on the topic \cite{chen2013unintentional, borodin2018big} suggesting that as the number of seats increases, the range of electoral outcomes decreases. However, for large values of $k$ we had to significantly downsample the branches in the sample tree that we explored given the exponential leverage. Therefore, the distribution markers are approximate, but the range is exact, as it was computed with the dynamic program on the full tree.

\begin{table*}
\centering
\begin{tabular}{llrrrrrrrrrrr}
\toprule
   &       &  $\ell$ &  \%infeas &  \%dup &  $H(b_i|b_j)$ &    ADS &  $\sigma_{50}$ &  $\sigma_{99}$ &  $\lambda_2$ &  $\mu_{walk}$ &    MCD &    ESR \\
State & $\epsilon_p$ &         &           &        &               &        &                &                &              &               &        &        \\
\midrule
IL & 0.005 &    2.81 &      0.34 &   1.17 &        0.0938 &  0.720 &          0.108 &          0.894 &        0.025 &         21.97 &  1.471 &  2.069 \\
   & 0.010 &    2.85 &      0.54 &   1.17 &        0.0936 &  0.711 &          0.108 &          0.893 &        0.024 &         22.60 &  1.472 &  2.312 \\
   & 0.025 &    2.79 &      0.32 &   1.15 &        0.0936 &  0.712 &          0.108 &          0.895 &        0.025 &         22.59 &  1.537 &  2.087 \\
   & 0.050 &    2.86 &      0.23 &   1.12 &        0.0929 &  0.717 &          0.108 &          0.893 &        0.024 &         22.15 &  1.561 &  2.159 \\
   & 0.100 &    2.82 &      0.21 &   0.94 &        0.0925 &  0.720 &          0.106 &          0.890 &        0.023 &         21.86 &  1.578 &  2.435 \\
NC & 0.005 &    1.53 &      0.12 &   2.86 &        0.1120 &  0.754 &          0.102 &          0.861 &        0.029 &         31.04 &  1.450 &  2.108 \\
   & 0.010 &    1.47 &      0.15 &   2.65 &        0.1116 &  0.752 &          0.103 &          0.859 &        0.031 &         30.98 &  1.352 &  1.977 \\
   & 0.025 &    1.48 &      0.06 &   2.53 &        0.1114 &  0.748 &          0.102 &          0.858 &        0.030 &         31.09 &  1.339 &  2.123 \\
   & 0.050 &    1.55 &      0.04 &   2.66 &        0.1093 &  0.752 &          0.101 &          0.857 &        0.027 &         30.80 &  1.360 &  2.127 \\
   & 0.100 &    1.47 &      0.03 &   2.21 &        0.1111 &  0.745 &          0.102 &          0.861 &        0.028 &         31.09 &  1.357 &  2.125 \\
\bottomrule
\end{tabular}
\caption{Generation results with varying population tolerance: leverage ($\ell$), percent infeasible of attempted partition problems (\%infeas), percent of districts that are duplicates (\%dup), the conditional binary entropy of two blocks appearing in the same district ($H(b_i|b_j)$), the average district similarity (ADS), the rank ratio of the block district matrix singular value need to reach the 50\% and 99\% reconstruction accuracy ($\sigma_{50}, \ \sigma_{99}$), the second eigenvalue of the normalized graph Laplacian with weights given by the block district matrix ($\lambda_2$), the average distance required to walk to the center of a district ($\mu_{walk}$), the maximum compactness disparity (MCD), and the expected seat range (ESR).}
\label{tab:pop_tol}
\end{table*}

\begin{table*}
\centering
\begin{tabular}{lllllrrrrrrl}
\toprule
   &     & $w$(root) &  $w$ & $\epsilon_p$ &  runtime (m) &  \%dups &  infeasible &  $\ell$-lb &  $\ell$ &  $\ell$-ub &   plans/s \\
state & $k$ &           &      &              &              &             &             &            &         &            &           \\
\midrule
NC & 5   &       100 &    5 &       0.0025 &        14.00 &       0.421 &       0.000 &     -1.376 &  -0.044 &      0.721 &  2.56E+00 \\
   & 10  &       200 &    4 &        0.005 &        71.93 &       1.662 &       0.169 &     -1.247 &   1.314 &      2.817 &  9.54E+01 \\
   & 13  &       150 &    4 &         0.01 &        69.15 &       2.705 &       0.082 &     -1.004 &   2.225 &      4.415 &  9.79E+02 \\
   & 20  &       100 &    3 &         0.02 &        49.13 &       1.834 &       0.101 &     -0.543 &   3.302 &      6.256 &  1.46E+04 \\
   & 40  &        60 &    3 &        0.025 &        71.71 &       4.567 &       0.190 &      1.216 &   9.171 &     15.172 &  1.88E+10 \\
   & 50  &        50 &    3 &         0.03 &        85.33 &       6.068 &       0.426 &      2.206 &  12.198 &     19.741 &  2.23E+13 \\
   & 60  &        40 &    3 &        0.035 &        74.48 &       7.048 &       0.507 &      3.305 &  15.185 &     24.418 &  2.47E+16 \\
   & 80  &        30 &  2.4 &         0.04 &        47.51 &       7.556 &       0.930 &      3.948 &  16.067 &     26.476 &  1.86E+17 \\
   & 100 &        50 &  2.1 &        0.045 &        62.57 &       7.670 &       1.365 &      4.566 &  17.227 &     28.491 &  2.74E+18 \\
   & 120 &        40 &    2 &         0.05 &        63.56 &       9.047 &       2.480 &      5.495 &  19.570 &     32.362 &  5.64E+20 \\
   & 140 &        30 &    2 &         0.05 &        73.57 &      10.837 &       4.594 &      6.923 &  23.334 &     38.305 &  2.53E+24 \\
IL & 5   &       100 &    5 &       0.0025 &        21.69 &       0.144 &       0.197 &     -1.319 &  -0.011 &      0.778 &  1.56E+00 \\
   & 10  &       200 &    4 &        0.005 &       109.28 &       1.535 &       0.125 &     -1.261 &   1.327 &      2.802 &  6.68E+01 \\
   & 18  &       150 &    3 &         0.01 &        91.96 &       1.191 &       0.496 &     -0.680 &   2.814 &      5.403 &  3.02E+03 \\
   & 30  &        80 &    3 &         0.02 &        84.90 &       2.581 &       0.380 &      0.299 &   6.234 &     10.677 &  1.30E+07 \\
   & 40  &        60 &    3 &        0.025 &        91.42 &       3.645 &       0.710 &      1.236 &   9.228 &     15.192 &  1.60E+10 \\
   & 50  &        50 &    3 &         0.03 &       106.76 &       4.414 &       1.131 &      2.220 &  12.165 &     19.754 &  1.60E+13 \\
   & 59  &        40 &    3 &        0.035 &       104.32 &       5.236 &       0.675 &      3.152 &  14.797 &     23.907 &  7.79E+15 \\
   & 80  &        30 &  2.4 &         0.04 &        62.51 &       5.667 &       1.936 &      3.984 &  16.202 &     26.511 &  1.78E+17 \\
   & 100 &        50 &  2.1 &        0.045 &       102.60 &       5.266 &       2.808 &      4.515 &  17.157 &     28.440 &  1.60E+18 \\
   & 118 &        40 &    2 &         0.05 &        97.17 &       5.675 &       3.481 &      5.353 &  19.394 &     31.769 &  2.41E+20 \\
   & 140 &        30 &    2 &         0.05 &       164.95 &       7.350 &       5.388 &      6.876 &  23.546 &     38.259 &  2.04E+24 \\
\bottomrule
\end{tabular}
\caption{Generation statistics with varying number of districts.}
\label{tab:vary_k_generation}
\end{table*}

\section{Ensemble Experiment} \label{ensemble_experiment}
The output of the first stage of our optimization pipeline is a set of districts that can be composed into a massive ensemble of distinct plans. These plans are not independent, and therefore are not an appropriate replacement for Markov chains in most statistical experiments. However, as we note in the main text, outside of rigorous outlier analysis, preserving independence is not as useful as simply having a massive sample size. Therefore, the purpose of our generation is twofold. First, it serves as a reusable input to our final optimization step, since we can use the districts for any objective function. The second is for ensemble analysis. 

 Because our ensemble is of exponential size, we cannot directly calculate most distribution level metrics (e.g., the 75th percentile of plan compactness). Therefore, we use a pruning algorithm to trim the tree to a manageable size to make a full enumeration tractable while preserving as much solution diversity as possible. Using the full enumeration, it becomes trivial to compute distribution summary statistics. Finally, to ensure we capture the full distributional range, we augment the pruned tree enumeration with the plans that minimize and maximize the metric we are analyzing, using our efficient dynamic programming approach. 

\paragraph{Subsampling} More specifically, for a sample tree storing the ensemble, we specify the target size of the number of plans we want to explicitly enumerate. We then start pruning each internal node's sampled partitions in the order of the capacity of the internal nodes. This is because nodes with smaller capacity inherently control the districting of a smaller area and therefore add much less to the overall diversity of the ensemble. We continue to prune until there is only one sample left, since we do not want to introduce infeasibility into any branch of the tree. After pruning each sample partition, we recompute the number of feasible solutions of each node using our recursive formula for $P(R, s)$. This is an efficient operation, since we only have to recompute the size of parent nodes of the pruned partition. After all nodes with capacity $s=currentPruneSize$ have only one sample left, we start pruning samples with capacity $s+1$ until we reach the desired number of distinct plans. See Algorithm~\ref{subsample} for pseudo-code.

\begin{algorithm2e*}
\SetAlgoLined
\KwResult{internalNodes}
currentPruneSize = 2\;
\While{solutionCount(root) $<$ targetSize}{
    pruneCandidates = [node for node in internalNodes if node.size() == currentPruneSize]\;
    \For{node in pruneCandidates}{
        \If{solutionCount(root) $\leq$ targetSize}{
            \KwRet\;
        }
        \If{node.children.length() $>$ 1}{
            node.children = node.children[:-1]\;
            recomputeNodeSolutionCount()\;
        }
    }
    \If{all([node.children.length() == 1 for node in internalNodes])}{
        currentPrunzeSize += 1\;
    }
}
\caption{Pruning Algorithm for Subsampling SHP Tree Ensemble}
\label{subsample}
\end{algorithm2e*}

\paragraph{Computing Compactness}
Most measures of compactness are actually very expensive to compute exactly because they require spatial aggregations of hundreds of polygons each consisting of hundreds or thousands of boundary points and then calculating pairwise distances for thousands of perimeter points of the resulting district. Additionally, some measures are only well-defined on a graph of atomic geographic blocks, but most states do not force atomicity of census tracts. In both cases, we require some approximations to make the calculations.

For the centralization measure of compactness, we compute the geographic centroid of each block ($x_i, y_i$). Then for each district, consisting of blocks $1, \dots, n$, we compute the population-weighted centroid as the weighted average of the block level centroids. So for a given district centroid ($x_{centroid}, y_{centroid}$), the overall district centralization is
$$\frac{1}{np}  \sum_{i=1}^n p_i \sqrt{(x_{centroid} - x_i)^2 + (y_{centroid} - y_i)^2}$$
which is approximately equal to the distance the average constituent would have to walk to reach the center of the district.

The Roeck measure of compactness is equal to the area of the district divided by the smallest circumscribing circle. Computing the diameter of a district exactly is a very expensive operation so instead we approximate the diameter by the maximum distance between any two block centroids. Since we only need to compute the area of a tract once, we get the exact area of the district using a simple sum. Therefore, in some cases, our Roeck measure has a compactness greater than one, but this is impossible in the exact case.

The cut-edges measure of compactness requires computing the number of edges that must be cut from the state adjacency graph $G$ to make a district. We do this exactly, however, for existing plans that are not made by composition of atomic blocks, we have to approximate the existing district. For each block (census tract), we compute its area overlap with each district, and assign it to the district with maximum overlap. Census tracts follow county boundaries, and often follow smaller municipal boundaries, and so in practice, the approximation is almost always very good if not exact.

\paragraph{Results}
For the ensemble presented in the main text, we use the sample fan-out width $w$ and number of root partitions $w(root)$ given by Table~\ref{tab:ensemble_statistics} and generate a large number of districts that admit an exponential ensemble of distinct district plans, also given by Table~\ref{tab:ensemble_statistics}. For all states, we used a population tolerance of $\epsilon_p = 0.01$. The center selection method was a combination of fixed-center and Pareto weight perturbation, where we fixed one center uniformly at random, applying a random Pareto perturbation (with $\alpha_{pareto} = 1$) to all of the weights, and then ran $k$-means (we refer to this method elsewhere as FC+1-P). Finally we used Voronoi weights with matched capacities. For selected metrics of interest, we compare our ensemble distribution with actual congressional elections occurring between 2012 and 2018. 

\begin{figure*}
    \includegraphics[width=\linewidth]{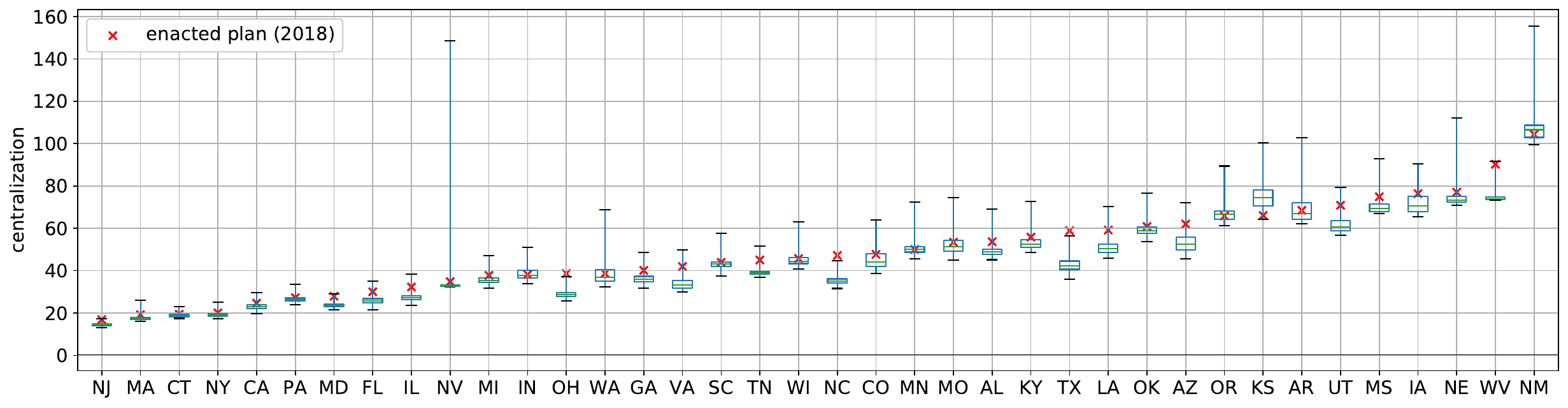}
    \includegraphics[width=\linewidth]{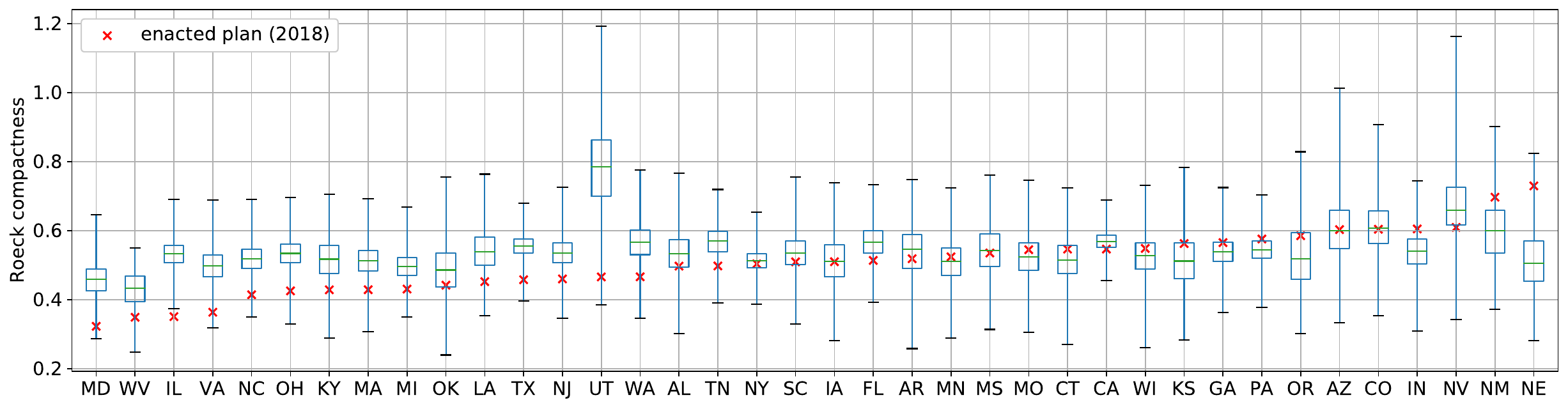}
    \includegraphics[width=\linewidth]{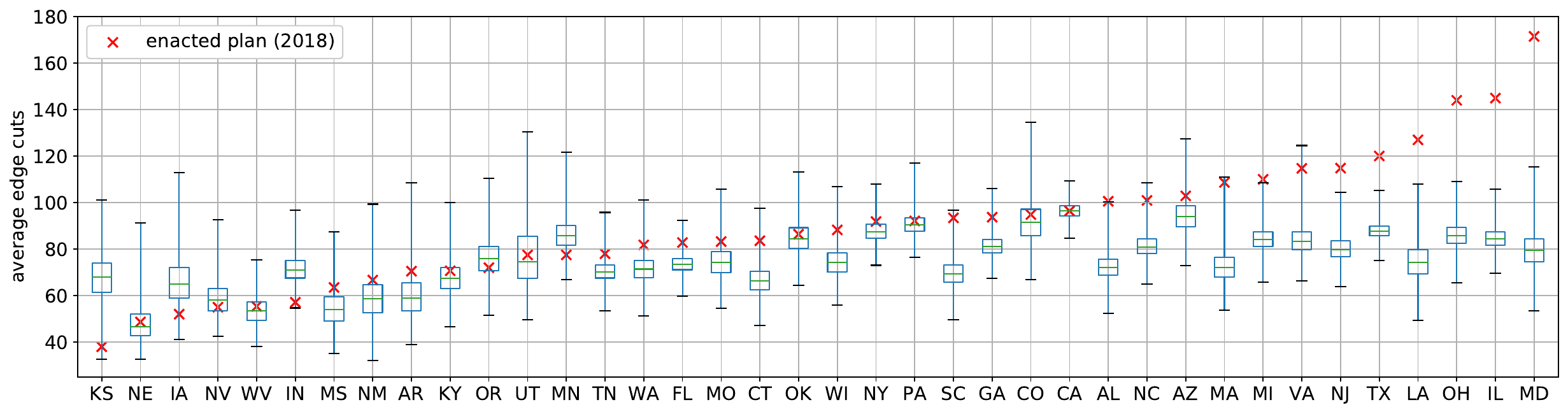}
    \caption{Distribution of plan compactness for all states with three or more seats compared with actual 116th House districts (2018). The box center indicates the median, with the top and bottom of the box indicating the 75th and 25th percentiles respectively of the down-sampled ensemble. The whiskers denote the minimum and maximum value of any plan found using the dynamic programming tree optimization algorithm.}
    \label{fig:compactness_distribution}
\end{figure*}

By construction, our algorithm only generated what we call \textit{natural} districts, which are districts with reasonable shapes that pass an eyeball test. To show this, we compare our ensemble with the actual districts from 2018 (see Figure~\ref{fig:compactness_distribution}). Cut-edges, the measure that we think does the best job of capturing the notion of compactness that is relevant for redistricting, shows how our method is incapable of producing contorted districts that are typical of gerrymandering. For instance, in Maryland and Ohio, two states well known for their gerrymandered districts, our least compact plan is far more compact than their enacted plan. Furthermore, in no state is there a plan that is more compact than our optimal one.

\begin{figure*}
    \includegraphics[width=\linewidth]{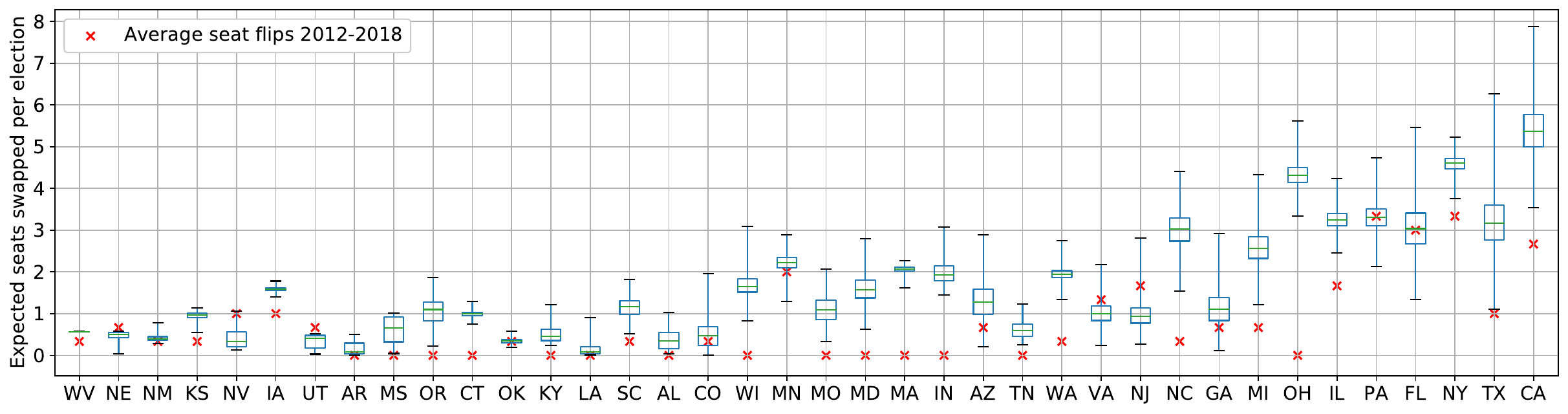}
    \caption{The expected number of partisan seat swaps ordered by total seats for all multi-district states. We include the average number of partisan seat swaps from the three transitions from 2012 to 2018. The box center indicates the median, with the top and bottom of the box indicating the 75th and 25th percentiles respectively of the down-sampled ensemble. The whiskers denote the minimum and maximum value of any plan found using the dynamic programming tree optimization algorithm.}
    \label{fig:seat_swap_distribution}
\end{figure*}

In Figure~\ref{fig:seat_swap_distribution}, we show the distribution of the expected number of partisan seat swaps for each election of our plan ensemble. The observed average of the actual enacted plan is much lower in almost every state than what is predicted by our ensemble. The reasons for this are twofold. First, many states engage in incumbent protection gerrymandering, as this is both less obvious and less controversial than more blatant partisan gerrymandering. Power dynamics more often encourage this type of gerrymandering because legislatures can often agree on keeping incumbents safe when they cannot accomplish partisan gerrymandering. Second, our affiliation model is not suited for this type of prediction because it does not take into account incumbency advantage, which in close congressional elections can be quite significant. Therefore, our model overstates the competitiveness of most elections.

Next, we discuss the results of the seat-share distribution, which we included in the main text. When the average observed seat-share from 2012 to 2018 falls outside of the generated ensemble, there are two main explanations. The first is gerrymandering occurred; that is, even when we tried to explicitly optimize for partisanship with natural districts, we could not achieve as extreme an outcome as the enacted plan. This is the case for Maryland, Virginia, Michigan, Pennsylvania, North Carolina, and Ohio. The other case is that statewide elections from the past decade do not generalize well to congressional elections. This is the case in New York and Massachusetts and are typically the result from incumbency advantage inflating one party's statewide share (as in New York) or a few popular moderate candidates from the party opposite of the traditional congressional delegation get elected (as in Massachusetts).

Cases where the observed average seat-share is both far from the ideal (with respect to the efficiency gap) but well within the ensemble range of expected outcomes typically indicate a natural gerrymander. These are cases where the enacted plan could have been drawn from random in an unbiased way, but the state's political geography allows for random plans that are highly unrepresentative. This is the case in Connecticut, Oregon, Colorado, Wisconsin, Florida, Georgia, Missouri, South Carolina, Louisiana, Nebraska, Alabama, and Arkansas. In most of these states, a plan exists that would be more fair, but typically requires the party in the majority to give up a seat to the minority party.

The political geography of a state has an enormous impact on the overall range of possible expected seat-shares. States dominated by one large city on the geographic exterior (like New York and Illinois) have smaller ranges because it is difficult to crack and pack these areas for an advantage. In other states, the background partisanship is high enough that it is difficult to construct any districts of the opposite party as is the case in Massachusetts, Connecticut, West Virginia, and Oklahoma. States with the largest ranges typically have one or more smaller cities in the geographic interior of the state which can be packed and cracked to greater effect like Colorado, Georgia, Arizona, and Missouri.

\begin{figure*}
    \includegraphics[width=\linewidth]{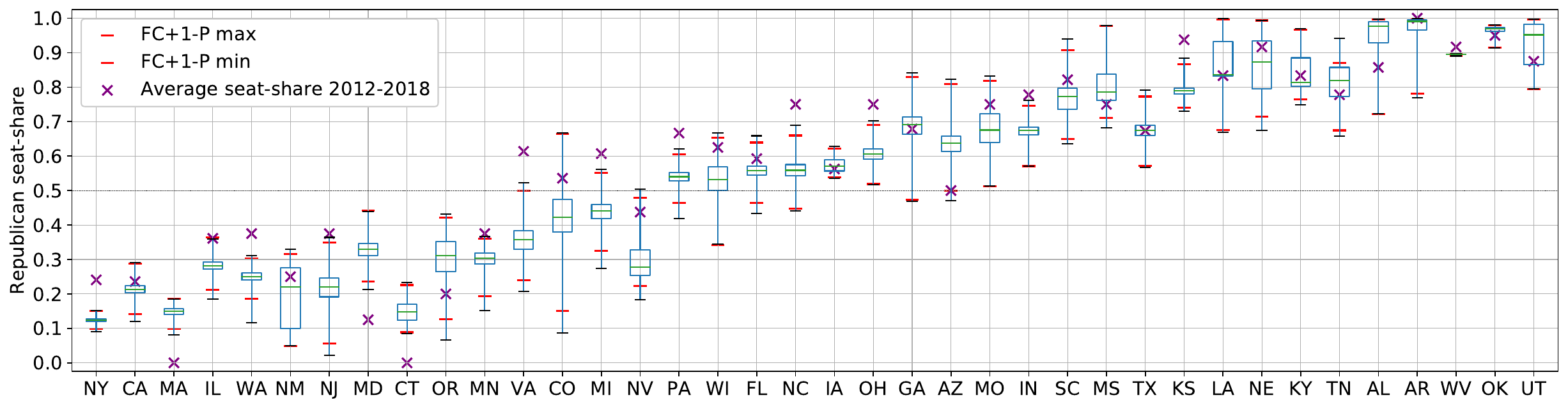}
    \caption{Comparison of ensemble distribution of expected seat share between uniform-random centers (boxplot indicating the min, quartiles, and max) and the range of the ensemble with the $k$-means based center selection method (FC+1-P) given in the main text.}
    \label{fig:seat_share_ensemble_comparison}
\end{figure*}

\begin{figure*}
    \includegraphics[width=\linewidth]{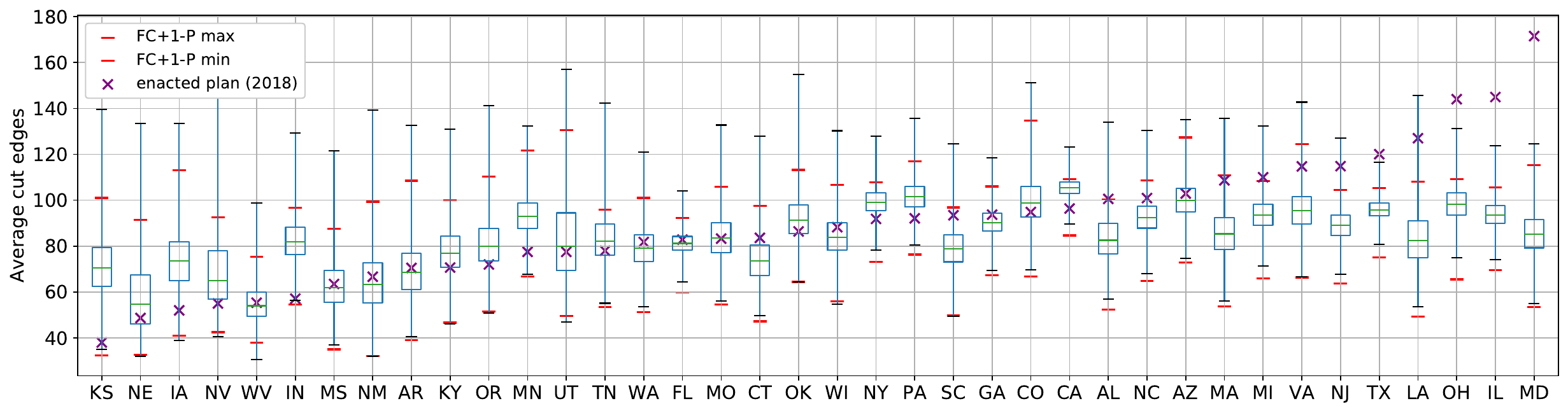}
    \caption{Comparison of ensemble distribution of plan compactness between uniform-random centers (boxplot indicating the min, quartiles, and max) and the range of the ensemble with the $k$-means based center selection method (FC+1-P) given in the main text.}
    \label{fig:compactness_ensemble_comparison}
\end{figure*}

\paragraph{Ensemble Comparison}
To test the sensitivity of our results, we ran the same ensembling experiment, except we used the uniform-random center selection method. The headline result is that compared to the ensemble presented in the main text (using 1 fixed random center with Pareto perturbation of the weights), in the extreme case, Democrats pick up 8 more seats in expectation and the Republicans 5 additional seats (see Figure~\ref{fig:seat_share_ensemble_comparison}). This makes the total achievable expected swing 22\% of all congressional seats. The cost of this additional swing, is compactness. Using completely random centers enables the creation of much more contorted districts, as compared to the $k$-means based center selection method (see Figure~\ref{fig:compactness_ensemble_comparison}).

\begin{table*}
    \centering
\begin{tabular}{lrrrrrrr}
\toprule
{} &  w(root) &   w &  generated districts &      generated plans &  leverage &  runtime &  subsampled plans \\
state &          &     &            &               &           &      (minutes)    &                   \\
\midrule
AL    &      885 &  10 &     216345 &  4.467369e+06 &      1.31 &   508.31 &             49000 \\
AR    &     1733 &  15 &     108054 &  1.013150e+05 &     -0.03 &   264.78 &             16000 \\
AZ    &      655 &   9 &     258302 &  5.108273e+07 &      2.30 &   701.10 &             81000 \\
CA    &       78 &   3 &      99910 &  1.483866e+18 &     13.17 &  1485.15 &           2809000 \\
CO    &      885 &  10 &     218721 &  4.651932e+06 &      1.33 &   650.02 &             49000 \\
CT    &     1326 &  13 &     153018 &  3.881580e+05 &      0.40 &   416.14 &             25000 \\
FL    &      175 &   4 &     154746 &  4.639542e+12 &      7.48 &   741.64 &            729000 \\
GA    &      385 &   6 &     215785 &  1.692028e+09 &      3.89 &   635.31 &            196000 \\
HI    &     3981 &  25 &       7962 &  3.981000e+03 &     -0.30 &    15.48 &              3981 \\
IA    &     1733 &  15 &     113136 &  1.102050e+05 &     -0.01 &   405.88 &             16000 \\
ID    &     3981 &  25 &       7962 &  3.981000e+03 &     -0.30 &    14.43 &              3981 \\
IL    &      285 &   5 &     200550 &  2.336721e+10 &      5.07 &   763.98 &            324000 \\
IN    &      655 &   9 &     277663 &  5.882950e+07 &      2.33 &   908.61 &             81000 \\
KS    &     1733 &  15 &     116790 &  1.032750e+05 &     -0.05 &   407.24 &             16000 \\
KY    &     1065 &  12 &     190497 &  1.853949e+06 &      0.99 &   688.58 &             36000 \\
LA    &     1065 &  12 &     196712 &  1.981538e+06 &      1.00 &   769.28 &             36000 \\
MA    &      655 &   9 &     275559 &  5.411175e+07 &      2.29 &  1089.84 &             81000 \\
MD    &      754 &  10 &     311335 &  2.603542e+07 &      1.92 &  1242.96 &             64000 \\
ME    &     3981 &  25 &       7962 &  3.981000e+03 &     -0.30 &    21.38 &              3981 \\
MI    &      385 &   6 &     220235 &  1.715810e+09 &      3.89 &  1053.72 &            196000 \\
MN    &      754 &  10 &     284717 &  2.286056e+07 &      1.90 &  1206.14 &             64000 \\
MO    &      754 &  10 &     275004 &  2.177805e+07 &      1.90 &  1231.05 &             64000 \\
MS    &     1733 &  15 &     111484 &  1.096030e+05 &     -0.01 &   448.07 &             16000 \\
NC    &      421 &   7 &     309805 &  2.485099e+09 &      3.90 &  1475.88 &            169000 \\
NE    &     2447 &  19 &      51333 &  2.444300e+04 &     -0.32 &   207.52 &              9000 \\
NH    &     3981 &  25 &       7962 &  3.981000e+03 &     -0.30 &    27.76 &              3981 \\
NJ    &      464 &   7 &     257116 &  6.118754e+08 &      3.38 &  1180.41 &            144000 \\
NM    &     2447 &  19 &      49569 &  2.356100e+04 &     -0.32 &   204.49 &              9000 \\
NV    &     1733 &  15 &     153288 &  1.465350e+05 &     -0.02 &   641.38 &             16000 \\
NY    &      175 &   4 &     246290 &  8.658624e+12 &      7.55 &  1420.58 &            729000 \\
OH    &      328 &   6 &     266918 &  2.161772e+10 &      4.91 &  1381.76 &            256000 \\
OK    &     1326 &  13 &     149034 &  3.827460e+05 &      0.41 &   728.31 &             25000 \\
OR    &     1326 &  13 &     141374 &  3.797200e+05 &      0.43 &   658.24 &             25000 \\
PA    &      285 &   5 &     198026 &  2.319985e+10 &      5.07 &  1085.40 &            324000 \\
RI    &     3981 &  25 &       7962 &  3.981000e+03 &     -0.30 &    31.11 &              3981 \\
SC    &      885 &  10 &     204303 &  4.408212e+06 &      1.33 &   940.35 &             49000 \\
TN    &      655 &   9 &     274535 &  5.854909e+07 &      2.33 &  1498.59 &             81000 \\
TX    &      124 &   3 &      87560 &  9.997533e+12 &      8.06 &   639.95 &           1296000 \\
UT    &     1733 &  15 &     105434 &  1.087480e+05 &      0.01 &   608.05 &             16000 \\
VA    &      515 &   8 &     300799 &  4.270091e+08 &      3.15 &  1927.57 &            121000 \\
WA    &      577 &   8 &     270498 &  1.215134e+08 &      2.65 &  1373.18 &            100000 \\
WI    &      754 &  10 &     262757 &  1.996886e+07 &      1.88 &  1542.86 &             64000 \\
WV    &     2447 &  19 &      51657 &  2.460500e+04 &     -0.32 &   261.71 &              9000 \\
\bottomrule
\end{tabular}
    \caption{State level ensemble experiment parameters and selected statistics. $w(root)$ is the number of sampled root partitions; $w$ is the samples per nonroot internal node. For each state the number of leaf nodes in the sample tree is the number of generated districts which admit a certain number of district plans. The leverage is the $\log_{10}(|P|/|D|)$, the log of the quotient of plans and districts. We report the runtime in minutes of generating the sample tree. Finally, to make calculating ensemble statistics tractable, we subsample to a smaller set of plans.}
    \label{tab:ensemble_statistics}
\end{table*}

\section{Empirical Runtime and Convergence} \label{runtime}

Finally, we present empirical results of the observed wall clock runtime of different parts of our algorithm and explore evidence of convergence. Given that our partitions are random, it is important we can sample a large number of feasible partition in a reasonable about of time. Each partition involves a number of steps: sampling centers, matching centers to capacities, computing the shortest-path tree for each center (for root partitions this is precomputed), constructing the optimization model, and finally solving the optimization model to optimality.

\begin{figure*}
    \centering
    \includegraphics[width=.495\linewidth]{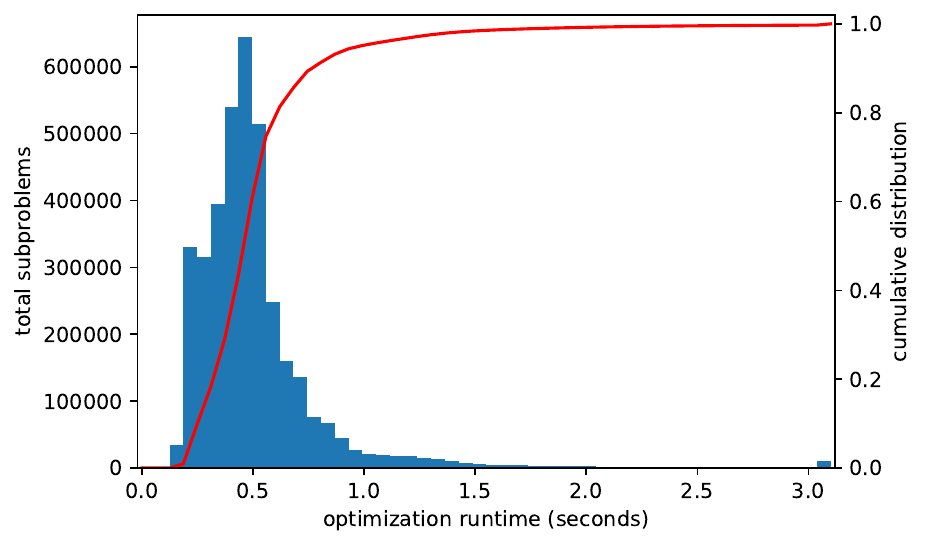}
    \includegraphics[width=.495\linewidth]{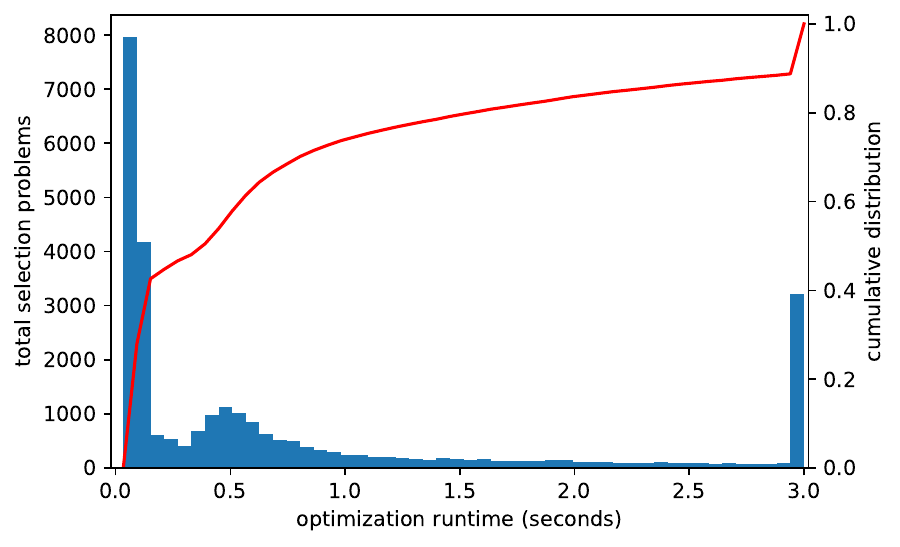}
    \caption{Histogram and cumulative distribution of the wall clock runtime to solve every partition problem (top) and master selection problem (bottom) in the 43 state ensemble presented in the main text. For illustration, the results were clipped at three seconds, but in both cases there is a long tail. For partition time, the runtime includes all required computational steps: center selection, capacity matching, computing the shortest path tree, model construction, and model optimization. For master selection problems, the total runtime includes model construction and model optimization to optimality.}
    \label{fig:runtime}
\end{figure*}

Figure~\ref{fig:runtime} plots the runtimes of every sampled partition in our main ensembling experiment. Of the 3715573 partition problems we solved, 62.1\% took less than 0.5 seconds, 95.2\% took less than 1.0 seconds, and 99.7\% took less than 3.0 seconds. However, there is a long tail associated with the root partitions of large states like California, Texas, and New York. The largest problem, a root partition of California with split size $z = 5$ (requiring over 40000 variables) took 38.9 seconds to construct and solve.

Figure~\ref{fig:runtime} also plots the histogram and cumulative distribution of the runtime of constructing and solving every master selection problem. We solve one master selection problem per root partition (excluding subtrees of height 1 where the split size of the root equals $k$, resulting in only one solution). Of the 28497 nontrivial column sets, 30.5\% were solved to optimality in under 0.1 seconds, 57.3\% in under 0.5 seconds, 73.9\% in under 1.0 seconds, and 89.1\% in under 3.0 seconds. There exists a similarly long tail, resulting from the largest trees, with the longest model taking 97.75 seconds to construct and solve.

For overall runtime, if executed serially, generating all 43 sample trees in their entirety would take 22.1 days and solving all master section problems would would take about 10 hours. However, our pipeline is arbitrarily parallelizable, so runtime becomes more a function of available computing resources than overall algorithmic efficiency. There is also an intermediate step between generation and optimization where we have to compute district-level metrics for millions of districts. For simple weighted averages of tract-level metrics, this takes on the order of an hour, but for more complicated metrics (like computing the number of cut edges of a district) this step can take an additional day. Having high leverage helps with this intermediate step. For instance, to compute linear plan-level metrics for an ensemble with a million plans generated from scratch would require $k10^6$ district-level computations because $\ell = \log(\frac{1}{k})$, but with $\ell = 2$ we only have to generate $10^4$ districts, yielding a $100k$ runtime savings in district-level metric computations. For larger states, where our leverage is naturally higher, this is a significant advantage.

\begin{figure*}
    \vspace{-3em}
    \includegraphics[width=\linewidth]{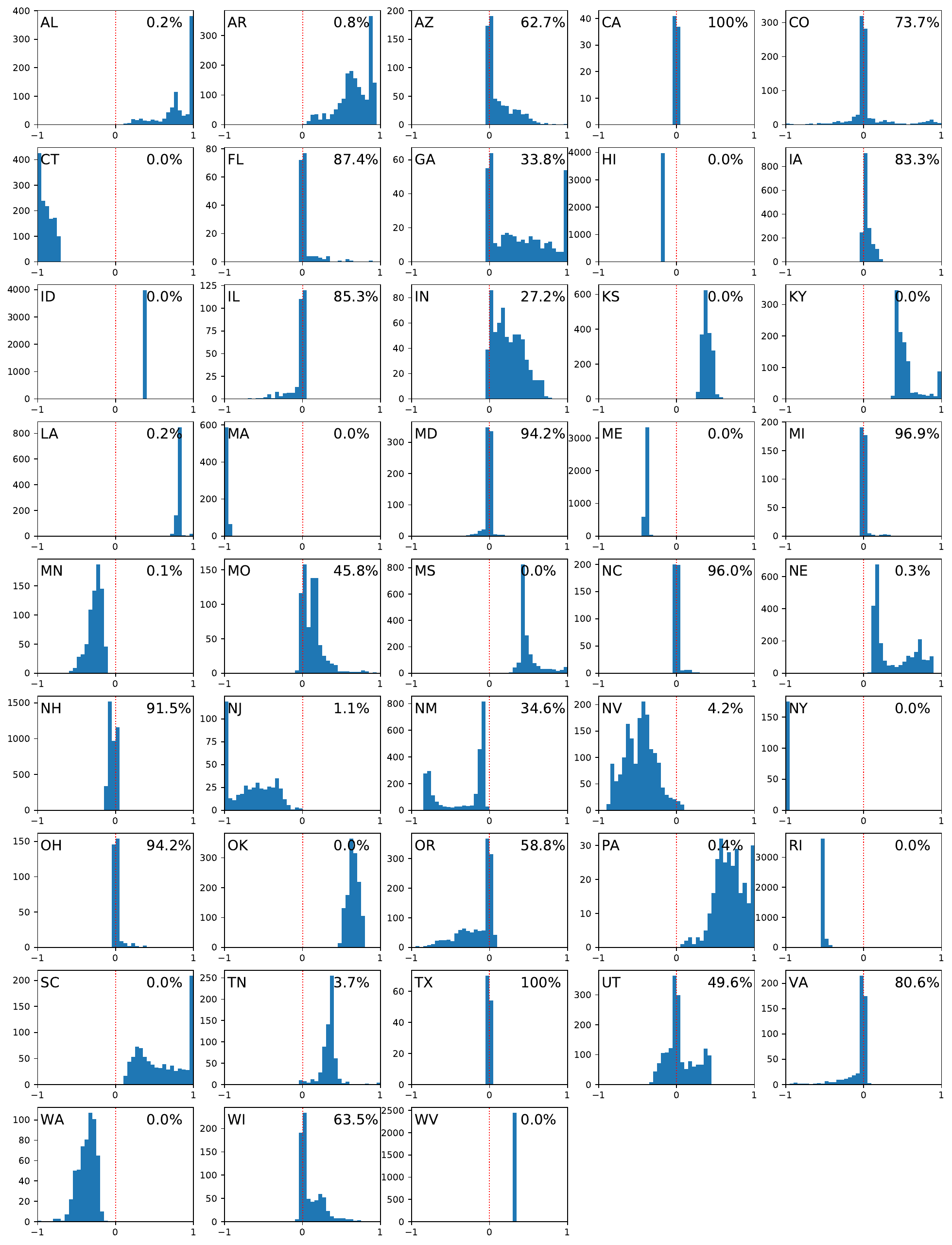}
    
    \vspace{-1em}
    \caption{Histogram of master selection problem optimal objective values for every root partition in 43 state ensemble. The $x$-axis gives the number of seats between the ideal number for a 0 efficiency gap solution and the expected seat total in the optimal solution, clipped to $\pm$ 1 seat. The upper right number gives the fraction of solutions $\pm 0.1$ seats from the optimal value.}
    \label{fig:convergence}
\end{figure*}

In Figure~\ref{fig:convergence}, we plot the histogram of master selection problem (MSP) optimal seat shares versus the ideal (0 estimated efficiency gap) for each root partition column set for each multi-district state. There are roughly three categories of states: those with an optimally fair solution for nearly every partition, those with solutions always far from fair, and those that have high variance yielding just a few root partitions with a fair solution. For instance, in Arizona, California, Colorado, Florida, Iowa, Illinois, Maryland, Michigan, North Carolina, New Hampshire, Ohio, Oregon, and Virginia over half of sampled root partitions yield a solution that is within 0.1 seats of the ideal expected seat share. In these states, for just the purpose of finding a fair solution, one iteration of root partitioning is likely enough. Of course, finding solutions that are fair and also compact, competitive, and maintain other state level rules would likely require more samples.

Then there are states like Connecticut, Hawaii, Idaho, Kansas, Kentucky, Massachusetts, Maine, New York, Rhode Island, and West Virginia, which have low MSP objective value variance with average objective value far from the optimally fair value. In these states, it is unlikely that any amount of root partition samples will yield a perfectly fair plan, without materially changing the constraints of the problem. Most of these states are simply too small to allow for finely tuned districts, and would require additional seats to more precisely calibrate outcomes.

Finally, there are states with high MSP objective value variance that rarely admit a 0 efficiency gap solution, or never do, but the variance is suggestive that such a solution could exist. This is the case in Alabama, Arkansas, Louisiana, Minnesota, Mississippi, Nebraska, New Jersey, Nevada, Pennsylvania, South Carolina, Tennessee, and Washington. A rich area of future work is to better quantify the convergence behavior of these methods to better direct computational effort.



\end{document}